\documentclass[letterpaper,titlepage,11pt]{article}

\usepackage{amssymb,amsmath,amsfonts}
\usepackage{epsfig}

\def\clock{{\count0=\time
           \divide\count0 60
           \ifnum\count0<10 0\fi\the\count0
           \multiply\count0 -60 \advance\count0 \time
           :\ifnum\count0<10 0\fi \the\count0
         }}
\newcommand{\timestamp}{{\small\vbox{\hbox{\tt\jobname.tex}
\hbox{\the\day/\the\month/\the\year, \clock}}}}

\newcommand{\CO}{\mathcal{O}}

\newcommand{\CM}{\mathcal{M}}

\newcommand{\BB}{\mathcal{B}}

\newcommand{\Z}{\mathbb{Z}}

\newcommand{\T}{\mathbb{T}}
\newcommand{\ms}{\mathfrak{s}}

\newcommand{\nn}{\nonumber}

\newcommand{\spa}{\ , \ \ }

\newcommand{\be}{\begin{equation}}
\newcommand{\ee}{\end{equation}}
\newcommand{\beq}{\begin{equation}}
\newcommand{\eeq}{\end{equation}}
\newcommand{\ben}{\begin{displaymath}}
\newcommand{\een}{\end{displaymath}}
\newcommand{\beqa}{\begin{eqnarray}}
\newcommand{\eeqa}{\end{eqnarray}}
\newcommand{\bea}{\begin{eqnarray}}
\newcommand{\eea}{\end{eqnarray}}
\newcommand{\bean}{\begin{eqnarray*}}
\newcommand{\eean}{\end{eqnarray*}}
\newcommand{\ba}{\begin{array}}
\newcommand{\ea}{\end{array}}
\newcommand{\bi}{\begin{itemize}}
\newcommand{\ei}{\end{itemize}}
\newcommand{\ie}{{\it i.e.,\,}}
\newcommand{\eg}{{\it e.g.,\,}}
\newcommand{\bbe}[1]{\mbox{${\mathbb E}^{#1}$}}
\newcommand{\bbr}[1]{\mbox{${\mathbb R}^{#1}$}}

\newcommand{\reef}[1]{(\ref{#1})}

\newcommand{\gsim}{\mathrel{\raisebox{-.6ex}{$\stackrel{\textstyle>}{\sim}$}}}
\newcommand{\lsim}{\mathrel{\raisebox{-.6ex}{$\stackrel{\textstyle<}{\sim}$}}}

\numberwithin{equation}{section}

\begin{document}

\begin{titlepage}
\begin{flushright}
CPHT-RR069.0707\\
\end{flushright}
\vskip 1.8cm
\begin{center}
{\bf\LARGE{The Phase Structure of Higher-Dimensional}} \vskip 0.12cm
{\bf\LARGE{Black Rings and Black Holes}} \vskip 1.6cm {\bf Roberto
Emparan$^{a,b}$, Troels Harmark$^{c}$,
Vasilis Niarchos$^{d}$,}\\
{\bf
Niels A. Obers$^{c}$,
Mar{\'\i}a J. Rodr{\'\i}guez$^{b}$
}
\vskip 0.5cm
\medskip
\textit{$^{a}$Instituci\'o Catalana de Recerca i Estudis
Avan\c cats (ICREA)}\\
\smallskip
\textit{$^{b}$Departament de F{\'\i}sica Fonamental, Universitat de
Barcelona, }\\
\textit{Diagonal 647, E-08028 Barcelona, Spain}\\
\smallskip
\textit{$^{c}$The Niels Bohr Institute,}
\textit{Blegdamsvej 17, 2100 Copenhagen \O, Denmark}\\
\smallskip
\textit{$^{d}$Centre de Physique Th\'eorique, \'Ecole Polytechnique,
 91128 Palaiseau, France}\\
 \textit{Unit\'e mixte de Recherche 7644, CNRS}

\vskip .2 in
{\tt emparan@ub.edu, harmark@nbi.dk, niarchos@cpht.polytechnique.fr,
obers@nbi.dk, majo@ffn.ub.es}
\end{center}
\vskip 0.3in

\baselineskip 16pt
\date{}

\begin{center}
{\bf Abstract}
\end{center}
\vskip 0.2cm \noindent We construct an approximate solution for an asymptotically flat,
neutral, thin rotating black ring in any dimension $D\geq 5$ by matching
the near-horizon solution for a bent boosted black string, to a
linearized gravity solution away from the horizon. The rotating black
ring solution has a regular horizon of topology $S^1 \times S^{D-3}$ and
incorporates the balancing condition of the ring as a zero-tension
condition. For $D=5$ our method reproduces the thin ring limit of the
exact black ring solution. For $D \geq 6$ we show that the black ring
has a higher entropy than the Myers-Perry black hole in the
ultra-spinning regime. By exploiting the correspondence between
ultra-spinning black holes and black membranes on a two-torus, we take
steps towards qualitatively completing the phase diagram of rotating
blackfolds with a single angular momentum. We are led to propose a
connection between MP black holes and black rings, and between MP black
holes and black Saturns, through merger transitions involving two kinds
of `pinched' black holes. More generally, the analogy suggests an
infinite number of pinched black holes of spherical topology
leading to a complicated pattern of connections and mergers between phases.

\end{titlepage} \vfill\eject

\setcounter{equation}{0}

\pagestyle{empty}
\small
\tableofcontents
\normalsize
\pagestyle{plain}
\setcounter{page}{1}

\newpage

\section{Introduction}

In this paper we explore the possible black hole solutions of the
Einstein equations $R_{\mu\nu}=0$ in six or more dimensions. Our
understanding of the black hole phases in five dimensions has
advanced greatly in recent years. In addition to the Myers-Perry
(MP) black holes \cite{Myers:1986un}, there exist rotating black
rings \cite{Emparan:2001wn,Emparan:2006mm} and multi-black hole
solutions like black-Saturns and multi-black rings
\cite{Elvang:2007rd,Elvang:2007hg,Iguchi:2007is,Evslin:2007fv}. The
latter have been constructed using inverse-scattering techniques
\cite{Belinsky:1971nt,Belinsky:1979,Belinski:2001ph,Pomeransky:2005sj},
which have also yielded a black ring solution and multi-black rings
with two independent angular momenta
\cite{Pomeransky:2006bd,biring}. There exist furthermore algebraic
classifications of spacetimes
\cite{DeSmet:2003rf,Coley:2004jv,Coley:2005ma,Pravda:2005qp},
theorems on how to determine uniquely the black hole solutions with
two symmetry axes
\cite{Morisawa:2004tc,Hollands:2007aj}%
\footnote{In \cite{Hollands:2007aj} it is proven that
five-dimensional stationary and axisymmetric solutions are unique
given the conserved asymptotic charges and the rod structure as
defined in \cite{Harmark:2004rm,Harmark:2005vn} (generalizing
\cite{Emparan:2001wk}).} and further solution generating techniques
\cite{Giusto:2007fx}. In fact, it is possible that essentially all
five-dimensional black holes (up to iterations of multi-black rings)
with two axial Killing vectors have been found by now. Parallel to
this progress, the phase diagram of black holes and black branes in
Kaluza-Klein (KK) spaces is also being mapped out with increasing
level of precision (see \textit{e.g.} the reviews
\cite{Kol:2004ww,Harmark:2007md} and the recent work
\cite{Dias:2007hg}).

In contrast to these advances, asymptotically flat vacuum solutions
with an event horizon in more than five dimensions remain largely a
\textit{terra incognita}, only known with certainty to be inhabited
by the MP solutions. Our aim is to begin to chart this landscape.
For reasons of simplicity, we shall confine ourselves to vacuum
solutions, $R_{\mu\nu}=0$, with angular momentum in only one of the
several independent rotation planes. However, our methods and many
of our conclusions should readily extend to more general situations.

One of the main results of this paper is the construction of an
approximate solution for an asymptotically flat, neutral, thin
rotating black ring in any dimension $D\geq 5$ with horizon topology
$S^1 \times S^{D-3}$. The method we will employ is the one of matched
asymptotic expansion
\cite{Harmark:2003yz,Gorbonos:2004uc,Karasik:2004ds,Gorbonos:2005px,Dias:2007hg}.
Our particular construction follows primarily the approach of
\cite{Harmark:2003yz}. First we find the metric of a thin black ring in
the linearized approximation to gravity sourced by the energy-momentum
tensor of an infinitely thin rotating ring. Then we obtain the
near-horizon metric by solving exactly the field equations to find
the perturbation of a boosted black string describing the bending of the
string into a circular shape. We finally match the two solutions in the
overlap zone, thereby completing the solution for a thin rotating black
ring. An important result of this exercise is that the perturbed event
horizon remains regular. As a further check of our method, we show that
in five dimensions our results for thin black rings are in agreement
with the thin ring limit of the exact rotating black ring metric.

In the process of constructing the black ring solution, we find that
the absence of naked singularities requires a zero-pressure
condition that corresponds to balancing the string tension against the
centrifugal repulsion. This is an example of how General Relativity
encodes the equations of motion of black holes as regularity
conditions on the geometry.

From the approximate rotating black ring solution we obtain the
asymptotic properties of the area function $\mathcal{A}(M,J)$ at
large spin and fixed mass, \textit{i.e.}\ at the ultra-spinning regime.%
\footnote{We find in particular that for $D \geq 6$ the leading form
of $\mathcal{A}(M,J)$ receives no corrections to first order in
$1/R$, where $R$ is the $S^1$ radius of the ring. On the contrary,
for $D=5$ there is an $1/R$ correction that matches
the thin ring limit of \cite{Emparan:2001wn}.} Remarkably, this
central result can be obtained with surprisingly little effort. The
limiting form of the function $\mathcal{A}(M,J)$ follows almost
immediately from the assumption that a rotating thin black ring is
well approximated as a boosted black string bent into a circle of
large radius $R$. The extra ingredient we need is the zero-pressure
condition that fixes the boost to a precise value.

From the area function $\mathcal{A}(M,J)$ for thin rotating black
rings we find that the entropy goes like
\begin{equation}
S (M,J) \propto J^{-\frac{1}{D-4}}\;M^{\frac{D-2}{D-4}}
\,,
\end{equation}
whereas for the ultra-spinning MP black holes in $D \geq 6$,
\begin{equation}
S (M,J) \propto J^{-\frac{2}{D-5}}\;M^{\frac{D-2}{D-5}}
\,.
\end{equation}
These results show that in the ultra-spinning regime of large $J$ for
fixed mass $M$ the rotating black ring has higher entropy than the MP
black hole.

The other main results of this paper concern the general phase diagram
of asymptotically flat neutral rotating black holes in six and higher
dimensions. To gain insight into this problem we exploit a connection
between, on one side, black holes and black branes in KK spacetimes and,
on the other side, higher-dimensional rotating black holes. This
connection was identified in \cite{Emparan:2003sy}, where it was argued
that a form of the Gregory-Laflamme instability of black branes
\cite{Gregory:1993vy,Harmark:2007md}, and the ensuing phases with
non-uniform horizons \cite{Gubser:2001ac,Wiseman:2002zc} (see also
\cite{Sorkin:2004qq,Kudoh:2004hs,Kleihaus:2006ee,Sorkin:2006wp}), arise
naturally in the ultra-spinning regime of black holes in six or more
dimensions. We can therefore use the known phase structure of the
solutions on a KK two-torus to take steps towards completing the phase diagram of
asymptotically flat rotating black holes. Then
the extension of the curves $\mathcal{A}(M,J)$ to all $J$ at a
qualitative level, and the inclusion of other new features of the phase
diagram, is a matter of well-motivated analogies, using the basic idea
in \cite{Emparan:2003sy}.

Employing this analogy, we develop further the conjecture in
\cite{Emparan:2003sy} that proposed the existence of `pinched' black
holes with spherical topology. We find a natural way to fit
them in the phase diagram, and connect them to black rings and black
Saturns through merger transitions.

In this paper, we do not claim to have the complete phase structure
of higher-dimensional black holes, even in the case with a single
rotation. Besides the uncertainties in our proposed completion of
known phase curves, there may be other black hole solutions (\eg
with other horizon topologies) beyond the MP solutions, the black
rings, and the pinched black holes. Nevertheless, the present work
is a first application of what we think is an effective approach to
learning about black holes in higher dimensions, at least at a
semi-quantitative level.

The outline of the paper is as follows. In section \ref{sec:thincircle}
we describe how viewing a black ring as a circular boosted black string
easily yields the area function $\mathcal{A}(M,J)$ in the ultra-spinning
regime. Section \ref{sec:matchasexp} outlines the perturbative
construction of thin black ring solutions.
The details are then
developed in sections \ref{sec:linear}, \ref{sec:equil} and
\ref{sec:nearhor}. In particular, in section \ref{sec:linear} we
construct the thin black ring metric in the asymptotic region where the
linearized approximation to gravity is valid. Then in section
\ref{sec:equil} we consider the overlap region between the asymptotic
region and the near-horizon region that enables us to derive the
zero-pressure equilibrium condition for thin black rings. Finally,
section \ref{sec:nearhor}, which is technically the most involved one,
considers the metric in the near-horizon region and proves that the
horizon of the black string can be bent and still remain regular.
Subsequently, section \ref{sec:phases} analyzes the resulting
thermodynamics for thin black rings, and compares them to MP
black holes. Section \ref{sec:complete} combines different pieces
of information to make a plausible conjecture for the qualitative
structure of the black hole phases in $D\geq 6$. Section
\ref{sec:discuss} discusses the results and outlook of this paper.
A number of appendices are also included. Appendix \ref{sec:flatcoords}
presents a set of coordinates in flat space that
are adapted to a circular ring.
Appendices \ref{app:fis}, \ref{app:regular} and \ref{app:FandGp} contain technical details
employed in sections \ref{sec:equil} and \ref{sec:nearhor}.
Appendix \ref{app:5D} shows how the known results for the
five-dimensional black rings are recovered from our method. Finally,
appendix \ref{app:torusphases} explains how to translate the results for
KK phases on $\T^2$ into a form appropriate for the correspondence to rotating black holes.

Readers who are mostly interested in the phase structure of
higher-dimensional black holes may skip the technical
construction of thin black ring solutions in sections \ref{sec:matchasexp} to
\ref{sec:nearhor}, and, after section \ref{sec:thincircle}, jump to
sections \ref{sec:phases} and \ref{sec:complete}.

Throughout this paper we shall denote the number of
spacetime dimensions $D$ via the number $n$ of  ``extra''
dimensions, \ie we set
\beq
D=4+n\,.
\eeq
This convention makes many equations a bit cleaner.

\section{Thin black rings from circular boosted black strings}
\label{sec:thincircle}

Black rings in $(n+4)$-dimensional asymptotically flat spacetime are
solutions of Einstein gravity with an event horizon of topology $S^1\times S^{n+1}$.
In five dimensions explicit solutions with
this topology have been presented in \cite{Emparan:2001wn}. However,
the construction of analogous solutions in more than five dimensions
is a considerably more involved
problem that has been unsuccessful so far---for instance,
 for $D \geq 6$ these solutions  are not contained in the  generalized
Weyl ansatz \cite{Emparan:2001wk,Harmark:2004rm,Harmark:2005vn}
since they do not have $D-2$ commuting Killing symmetries;
furthermore the inverse scattering techniques of
\cite{Belinsky:1971nt,Belinsky:1979,Belinski:2001ph,Pomeransky:2005sj}
do not extend to the asymptotically flat case in any $D\geq 6$. In
what follows, we will make progress towards solving this problem by
constructing thin black ring solutions in arbitrary dimensions in a
perturbative expansion around circular boosted black strings.

The metric of the \textit{straight} boosted black string is
\begin{eqnarray}
\label{appab}
ds^2 &=& -\left( 1 - \cosh^2 \alpha \frac{r_0^{n}}{r^{n}} \right) dt^2 -
 2 \frac{r_0^{n}}{r^{n}}
 \cosh \alpha \sinh \alpha\, dt dz + \left( 1 + \sinh^2 \alpha
\frac{r_0^{n}}{r^{n}} \right) dz^2 \nn \\ &&
 + \left( 1- \frac{r_0^{n}}{r^{n}} \right)^{-1} dr^2 + r^2 d\Omega_{n+1}^2 \,,
\end{eqnarray}
where $r_0$ is the horizon radius and $\alpha$ is the boost parameter.
In general, we will take the $z$ direction to be along an $S^1$ with
circumference $2\pi R$, which means we can write $z$ in terms of an
angular coordinate $\psi$ defined by
\beq
\psi=\frac{z}{R}\,,\qquad 0\leq \psi <2\pi\,.
\eeq
At distances $r\ll R$, the solution \eqref{appab} is the approximate
metric of a thin black ring to zeroth order in $1/R$.

By definition, a thin black ring has an $S^1$
radius $R$ that is much larger than its $S^{n+1}$ radius $r_0$. In this limit,
the mass of the black ring is small and the gravitational attraction
between diametrically opposite points of the ring is very weak. So,
in regions away from the black ring, the linearized approximation to
gravity will be valid, and the metric will be well-approximated if we
substitute the ring by an appropriate delta-like distributional source
of energy-momentum. The source has to be chosen so that the metric it produces
is the same as that expected from the full exact solution in the region
far away from the ring. Since the thin black ring is expected to
approach locally the solution for a boosted black string, it is sensible
to choose distributional sources that reproduce the metric \reef{appab}
in the weak-field regime,
\begin{subequations}
\label{distsource}
\beqa\label{distsourcea}
T_{tt}&=&\frac{r_0^{n}}{16\pi G}\,\left(n\cosh^2\alpha+1\right)\,\delta^{(n+2)}(r)\,,\\
\label{distsourceb}
T_{tz}&=&\frac{r_0^{n}}{16\pi G}\,n\cosh\alpha\sinh\alpha\,\delta^{(n+2)}(r)\,,\\
\label{distsourcec}
T_{zz}&=&\frac{r_0^{n}}{16\pi G}\,\left(n\sinh^2\alpha-1\right)\,\delta^{(n+2)}(r)\,.
\eeqa
\end{subequations}
The location $r=0$ corresponds to
a circle of radius $R$ in the $(n+3)$-dimensional Euclidean flat space,
parametrized by the angular coordinate $\psi$. We shall be more specific
about these coordinates in later sections.

In this construction the mass and angular momentum of the
black ring are obtained by integrating the energy and momentum densities,
\begin{subequations}\label{MJ}
\bea
\label{MJa}
M&=&2\pi R \int_{\BB^{n+2}} T_{tt}\, , \\
\label{MJb}
J&=&2\pi R^2 \int_{\BB^{n+2}} T_{tz}\,,
\eea
\end{subequations}
where $\BB^{n+2}$ intersects the ring once so
$\frac{1}{\Omega_{n+1}}\int_{\BB^{n+2}}\delta^{(n+2)}(r)=1$. Moreover, the area of the black
ring is that of a
boosted black string of length $2\pi R$,
\beq\label{area}
\mathcal{A}=2\pi R\,\Omega_{n+1}\,  r_0^{n+1}\cosh\alpha\,.
\eeq

Since $g_{tz} + \tanh
\alpha\, g_{zz} = 0$ at the horizon $r=r_0$, the linear
velocity of the horizon is $v_H=\tanh\alpha$, hence
the angular velocity of the black ring in the $\psi$-direction is
\beq
\label{appac}
\Omega_H=\frac{v_H}{R}=\frac{\tanh \alpha}{R}\,.
\eeq
The surface gravity is the same as that of a boosted black string,
\beq\label{kappa0}
\kappa=\frac{n}{2 r_0\cosh\alpha}\,.
\eeq

Such a black ring is described by three parameters:
$r_0$, $R$ and $\alpha$ (see Refs.~\cite{Hovdebo:2006jy,Kastor:2007wr}
for further details on boosted black strings and their thermodynamics).
More physically, we can take the parameters to be
$M$, $J$, $R$, and so in principle the area is a function
$\mathcal{A}(M,J,R)$. However, a black ring in mechanical equilibrium
should be characterized by only two
parameters: given, say, the mass and the radius, there should be only
one value\footnote{Or possibly a discrete number of them, but this
requires large self-gravitational effects, beyond the reach of our linearized
approximation.} of the angular momentum
for which the ring is in equilibrium.

It is actually easy to find the dynamical balance condition that
relates the three parameters. We are approximating the black ring by a
distributional source of energy-momentum. The general form of the
equation of motion for probe brane-like objects has been determined in
\cite{Carter:2000wv}. In the absence of external forces it takes the form
\beq\label{KT}
{K_{\mu\nu}}^{\rho}T^{\mu\nu}=0\,,
\eeq
where the indices $\mu,\nu$ are tangent to the brane and $\rho$ is
transverse to it. The second fundamental tensor
${K_{\mu\nu}}^{\rho}$ extends the
notion of extrinsic curvature to submanifolds of codimension possibly larger than
one. The extrinsic curvature of the circle is $1/R$, so a circular
linear distribution of energy-momentum of
radius $R$ will be in equilibrium only if
\beq\label{notzz}
\frac{T_{zz}}{R}=0\,,
\eeq
\ie for finite radius the pressure tangential to the circle must
vanish. Hence, for our thin black ring with source \reef{distsource}, the
condition that the ring be in equilibrium translates into a very specific value
for the boost parameter
\beq\label{eqboost}
\sinh^2\alpha=\frac{1}{n}\,.
\eeq
Then, eqs.~\reef{MJ} and \reef{area} become
\begin{subequations}\label{MJA}
\beqa\label{MJAa}
M&=&\frac{\Omega_{n+1}}{8 G}\,R\, r_0^{n}(n+2)\,, \\
\label{MJAb}
J&=&\frac{\Omega_{n+1}}{8 G}\,R^2\, r_0^{n}\sqrt{n+1}\,,\\
\label{MJAc}
\mathcal{A}&=&\Omega_{n+1}\, 2\pi
R\,r_0^{n+1}\sqrt{\frac{n+1}{n}}\,.
\eeqa
\end{subequations}
Notice that an equivalent but more physical form of the
equilibrium equation \reef{eqboost} is
\beq
\label{RJM}
R=\frac{n+2}{\sqrt{n+1}}\frac{J}{M}\,.
\eeq
We see that the radius grows linearly with $J$ for fixed mass.

In principle we can eliminate $r_0$ and $R$ from \reef{MJAc} to express
$\mathcal{A}$ as a function of $M$ and $J$. This relation is one of the
main results in this paper and it will be studied in detail later in
section~\ref{sec:phases}. But before that, there are two issues in our
analysis that demand further scrutiny.

First, the above reasoning relies crucially on the assumption that when
the boosted black string is curved, the horizon remains regular. To
verify this point, and also to obtain a metric for the thin black ring,
we shall solve the equations and construct an approximate solution for
$r_0\ll R$ using a matched asymptotic expansion. This is done in
sections \ref{sec:matchasexp} to \ref{sec:nearhor}.

Second, the black ring must be a solution to the source-free Einstein
vacuum equations, and so the point may be raised that our derivation of
the equilibrium condition \reef{notzz} is based only on the
properties of (distributional) sources. After all, in the exact
five-dimensional black ring solution of \cite{Emparan:2001wn}, the equilibrium
condition was not derived from any equation such as \reef{KT}, but
instead by demanding the absence of singularities
on the plane of the ring outside the horizon.
We will see that the use of a matched asymptotic expansion
allows us to produce a similar proof of \reef{notzz}: whenever
$n\sinh^2\alpha\neq 1$ with finite $R$, the geometry backreacts
creating singularities on the plane of the ring. These
singularities admit a natural interpretation. Since \reef{KT} is a
consequence of the conservation of the energy-momentum tensor, when
\reef{notzz} is not satisfied there must be additional sources of
energy-momentum. These additional sources are responsible
for the singularities in the geometry.
Alternatively, the derivation of \reef{notzz} appearing below is an
example of how General Relativity encodes the equations of motion of
black holes as regularity conditions on the geometry.

The idea that thin black rings should be well approximated by boosted
black strings is certainly not new. The black string limit of
five-dimensional black rings was first made explicit in
\cite{Elvang:2003mj}, where the condition $T_{zz}=0$ at equilibrium was
also noticed---but the connection with \reef{KT} was missed. An attempt
to describe thin black rings in higher dimensions with the use of
boosted black strings appeared in \cite{Hovdebo:2006jy}, where the
authors correctly predicted the existence of black rings in $D\geq 6$.
They also attempted to determine the equilibrium boost of the black
string by maximizing the entropy with respect to the boost. This
approach is clearly different from the one here, which is instead based
on mechanical equilibrium, and leads to incorrect dynamics. Even in the
limit of very large
number of dimensions, where the results of
\cite{Hovdebo:2006jy} simplify, maximization of the entropy predicts
$\frac{J}{MR} \to \frac{1}{\sqrt{2n}}$ instead of the correct
$\frac{J}{MR} \to \frac{1}{\sqrt{n}}$ (see eq.\ \eqref{RJM}). The
mechanical viewpoint was advocated in \cite{Elvang:2006dd}, where a
simple Newtonian model was shown to reproduce the correct equilibrium in
5D and so ref.~\cite{Elvang:2006dd} independently pointed out the
possibility of equilibrium configurations of thin rotating black rings
in all $D\geq 5$. However, the model does not capture correctly all the
relativistic ring dynamics, and would predict a value of the boost in
conflict with $T_{zz}=0$ in $D\geq 6$, so the authors of
\cite{Elvang:2006dd} chose not to make any quantitative predictions.

\section{Matched asymptotic expansions}
\label{sec:matchasexp}

Our aim is to construct approximate solutions for thin black rings
following a systematic approach that, in principle, allows to proceed
iteratively by starting from a limit where the solution is known and
then correcting it in a perturbative expansion. The method was first
used in \cite{Harmark:2003yz} to construct black holes localized on a KK circle,
and then subsequently refined and extended in the same context in
\cite{Gorbonos:2004uc,Karasik:2004ds,Gorbonos:2005px,Dias:2007hg}.
Ref.~\cite{Gorbonos:2004uc}
provides a lucid description of the method.

The basic idea is that, in a problem with two widely separated scales,
we can try to find approximate solutions to the equations in two zones
and match them at an intermediate zone where both approximations are
valid. In our problem, the two scales are $r_0$ and $R$. We will first
consider an asymptotic zone at large distances from the black ring,
$r\gg r_0$, where the field can be expanded in powers of $r_0$. Next
we will consider the near-horizon zone that lies at scales much smaller than the
ring radius, $r\ll R$. In this zone the field is expanded in powers of $1/R$. At
each step, the solution in one of the zones is used to provide boundary
conditions for the field in the other zone, by matching the fields in
the `overlap' zone $r_0\ll r \ll R$ where both expansions are valid.

In more detail, the first few steps in this construction, which are the
ones that we develop in this paper, proceed as follows:

\begin{enumerate}

\item[$\it 0$.] In the zeroth order step, we consider the solution in the
near-horizon zone to zeroth order in $1/R$, \ie we take a boosted black string
of infinite length, $R\to\infty$. Relating the parameters $r_0$, $R$ of
the black string to the asymptotic mass and angular momentum of the
black ring is most simply done using the analogue of Gauss' law in
General Relativity, namely, Stokes' theorem applied to the Komar
integrals for $M$ and $J$.

\item[$\it 1$.] Then we solve the Einstein equations in the linearized
approximation around flat space, for a source of the right form --- in
this case a circular distribution of a given mass and momentum density.
The linearized approximation is an expansion to first order in
$r_0^{n}\propto GM/R$, valid for $r_0\ll r$. The linearized solution is completely
determined (up to gauge transformations) by the sources and the boundary
conditions of asymptotic flatness.

\item[$\it 2.$] Next we focus on the near-horizon region of the ring.
The goal is to find the linear corrections to the metric of a boosted black
string for a perturbation that is small in $1/R$; in other words, we
analyze the geometry of a boosted black
string that is now slightly curved into a circular shape. To find these corrections,
one must solve a set of homogeneous equations, and therefore boundary
conditions must be provided. The matching to the solution of the
previous step in the overlap region $r_0\ll r\ll R$ provides boundary conditions
at large $r$ (with due care to the gauge choices in the two regions).
In addition, one must also pay attention to the regularity conditions at the
horizon $r\to r_0$.

\item[$\it 3.$] The near-horizon metric can now be used to fix the
integration constants that appear when one solves for the
next-to-linearized order corrections at large distances. We shall not
solve this step, since, as in step 0, there is always a simpler way to
obtain the physical asymptotic magnitudes $M$ and $J$ to this order
\cite{Harmark:2003yz}: the corrections to the metric near the horizon determine
the corrections to the area $\cal A$, temperature $T$, and angular
velocity $\Omega_H$. By using the Smarr relations and the first law, we
can immediately deduce the corrections to the mass and angular momentum.
In principle, these are measured at infinity, but the Smarr relations
and the first law can give the required information since they use
implicitly the equations of motion.

\end{enumerate}

Step 0 has been discussed already in section~\ref{sec:thincircle}. The
relation $\mathcal{A}(M,J)$ to leading order, following from \reef{MJA},
gives the asymptotic form of the phase curve of thin black rings. Simple
as this is, it nevertheless gives us non-trivial information, as it
tells us \eg whether MP black holes or black rings dominate the entropy
at large spins. Step 1 will be solved in sections \ref{sec:linear} and
\ref{sec:equil}. Step 2 is then developed in section \ref{sec:nearhor}.
Step 3, discussed at the end of section \ref{sec:nearhor}, gives the
leading order corrections to the entropy curve ${\cal A}(M,J)$. We will
actually find that these corrections vanish for thin black rings in six
or more dimensions.

\section{Black rings in linearized gravity}
\label{sec:linear}

We want to find the solution that describes a thin black ring, with
$S^{n+1}$ radius $r_0$ much smaller than its $S^1$ radius $R$, in the region
$r\gg r_0$ where the linearized approximation to gravity should be
valid.

We solve the linearized Einstein equations in transverse gauge,
\beq\label{eins}
\Box \bar h_{\mu\nu}=-16\pi G T_{\mu\nu}\,
\eeq
with $\bar h_{\mu\nu}=h_{\mu\nu}-
\frac{1}{2} h g_{\mu\nu}$ and $\nabla_\mu \bar h^{\mu\nu}=0$. Writing
the $(n+3)$-dimensional Euclidean flat space metric in bi-polar
coordinates
\beq\label{bipolar}
ds^2(\bbe{n+3})=dr_1^2+r_1^2 d\Omega^2_{n}+dr_2^2+r_2^2 d\psi^2
\eeq
we take the equivalent ring source to lie at $r_1=0$ and $r_2=R$.
Locally the source must reproduce a boosted black string, with non-zero
energy density $T_{tt}$ and angular momentum density $T_{t\psi}=R\,
T_{tz}$ as in \reef{distsource}.
Finding the solution valid at all $r_1$ and
$r_2$ with non-zero $T_{\psi\psi}=R^2\, T_{zz}$ is not easy, so we
shall assume that the equilibrium condition $T_{\psi\psi}=0$ is satisfied. In
this case the general form of the metric is
\beq\label{linsol}
ds^2=(-1+2\Phi)dt^2-2Adt d\psi+\left(1+\frac{2}{n+1}\Phi\right)ds^2(\bbe{n+3})\,,
\eeq
where $\Phi$ and $A$ depend only on $r_1,r_2$ and are sourced
respectively by $T_{tt}$ and $T_{t\psi}$. Their equations are
actually very simple: away from the source, $\Phi$ must solve the
scalar Laplace equation, and $A$ the Maxwell equation for a gauge potential
$A_\mu dx^\mu =A d\psi$. The solutions for the appropriate
distributional sources have
already been constructed in \cite{Emparan:2001ux} (see \cite{Lunin:2002iz} for further details).
For the scalar potential we have
\beq\label{Phisol}
\Phi= \frac{4 G M}{(n+2)\Omega_{n+2}}\int_{0}^{2\pi}d\psi\,
\frac{1}{\left(r_1^2 + (R\cos \psi - r_2)^2 + R^2 \sin^2
\psi \right)^{(n+1)/2}}
\eeq
and for the vector potential $A$,
\beq\label{Asol}
A= \frac{8 G J}{(n+1)\Omega_{n+2}R}\int_{0}^{2\pi}d\psi\,\frac{
r_2\cos\psi}{\left(r_1^2 +(R\cos \psi - r_2)^2 + R^2 \sin^2
\psi \right)^{(n+1)/2}}\,.
\eeq
These integrals take simple forms only when $n$ is odd \cite{Emparan:2001ux}, while
for even $n$ they involve elliptic functions \cite{Lunin:2002iz}. Nevertheless,
they can be approximately calculated both in the asymptotic region
$r_1,r_2\gg R$ and in the `overlap' zone $r_1,\,r_2- R\ll R$. The values
at asymptotic infinity
\beq
\Phi\to\frac{8\pi G M}{(n+2)\Omega_{n+2}}
\frac{1}{\left(r_1^2+r_2^2\right)^{\frac{n+1}{2}}}
\,,\qquad
A\to\frac{8\pi G J}{\Omega_{n+2}}\frac{r_2^2}{\left(r_1^2+r_2^2\right)^{\frac{n+3}{2}}}
\eeq
have been used to normalize the potentials in
terms of the ADM mass and angular momentum,
$M$ and $J$. Since the latter can also be computed
using Komar integrals, if we move the integration surface from
infinity towards the vicinity of the source \reef{distsource}, the relations
\reef{MJA} follow easily.

\section{The overlap zone: deriving the zero-tension condition}
\label{sec:equil}

In the previous section we found the linearized solution with
arbitrary mass and angular momentum, and zero tension. Finding the
solution with a source for tension is more complicated, but the task
becomes much easier if one restricts to the overlap zone $r_0\ll
r_1,\,r_2- R\ll R$. In this regime we are studying the
effects of locally curving a thin black string into an arc of constant
curvature radius $R$. We shall prove that a regular solution is possible
only if $T_{\psi\psi}= 0$.

\subsection{Adapted coordinates}
\label{sec:adapted}

In order to study the effect of curving the string into a circle of
radius $R$, it is very convenient to use adapted coordinates, \ie
coordinates that correspond to equipotential surfaces of the field of a
circular source. One can work out a system of such coordinates
valid both in the asymptotic and near-ring zones, and this
was indeed the approach taken in \cite{Harmark:2003yz} using the
coordinate system and ansatz introduced in \cite{Harmark:2002tr}.
While a similar exercise
can be carried out as well for this problem, see appendix
\ref{sec:flatcoords}, it is technically quite involved and in general
impractical. A simpler method goes as follows.

We work directly in the region $r\ll R$, to leading order in $1/R$.
In order to find coordinates for flat space so that $r=0$ is an
arc of ring of radius $R$, we seek a metric such that:
\begin{itemize}

\item The metric is Riemann-flat to order $1/R$.

\item The curve $r=0$ on the ring plane $\theta=0,\pi$ has constant extrinsic
curvature radius $R$.

\end{itemize}
We will also require that surfaces of constant radial coordinate $r$ are
equipotential surfaces of the Laplace equation\footnote{Another
possibility is to adapt $r$ to the potential of a three-form field
strength \cite{Emparan:2006mm}. However, this appears to be useful only for
$n=1$.} for a delta-function source at $r=0$:
\begin{itemize}

\item $\nabla^2 r^{-{n}}=0$.

\end{itemize}
Let us then make the ansatz
\beq
ds^2=\left(1+A(\theta)\frac{r}{R}\right)dz^2+
\left(1+B(\theta)\frac{r}{R}\right)dr^2+
\left(1+C(\theta)\frac{r}{R}\right)r^2(d\theta^2+\sin^2\theta
d\Omega_{n}^2)\,.
\eeq
Riemann flatness requires that $A(\theta)=A \cos\theta$, with
constant $A$. Also, up to irrelevant additive constants, it requires
$B(\theta)=C(\theta)=B \cos\theta$. Then, $\nabla^2 r^{-n}=0$ implies $A=-n
B$. The correct curvature radius $R$ is obtained for $A=2$.

Thus, the flat space metric, to first order in $1/R$, is
\begin{equation}\label{adapted}
ds^2(\bbe{n+3}) =\left( 1 + \frac{2r\cos \theta}{R}
\right) dz^2 +\left( 1 - \frac{2}{n}\frac{r \cos \theta}{R} \right)
\left( dr^2 + r^2 d\theta^2 + r^2 \sin^2 \theta d\Omega_{n}^2
\right)\, .
\end{equation}
The relation between this metric and \reef{bipolar} is studied in appendix
\ref{sec:flatcoords}.

In these coordinates, we consider a general distributional linear source of
mass, momentum and tension,
\begin{subequations}\label{Tmunu}
\beqa
T_{tt}&=&\frac{n(n+2)}{n+1}\;\mu\;\frac{r_0^{n}}{16\pi G} \;\delta^{(n+2)}(r) \label{Ttt} \,,\\
T_{tz}&=&n\,p\,\frac{r_0^{n}}{16\pi G}\; \delta^{(n+2)}(r)\,,\label{Ttz}\\
T_{zz}&=&\frac{n(n+2)}{n+1}\;\tau\;\frac{r_0^{n}}{16\pi G}\; \label{Tzz} \delta^{(n+2)}(r)\,.
\eeqa
\end{subequations}
This is more general than \reef{distsource}, since we allow the three
components to be independent. The source is parametrized with
three dimensionless quantities
$\mu,\,p,\,\tau$ (conveniently normalized to simplify later results) but
one of them could be absorbed in $r_0$. The
boosted black string has
\begin{subequations}\label{mutaup}
\beqa
\frac{n(n+2)}{n+1}\mu&=&n\,\cosh^2\alpha +1\,,\label{mubeta}\\
\frac{n(n+2)}{n+1}\tau&=&n\,\sinh^2\alpha -1\,,\label{taubeta}\\
p&=&\cosh\alpha\sinh\alpha\,. \label{pbeta}
\eeqa
\end{subequations}
In order to be more general, and to have a clearer notation, we choose
to work with a general source and with a superfluous parameter.

\subsection{Solving the equations}
\label{sec:solving}

In the adapted coordinates \reef{adapted}, using the symmetries of the
solution and appropriate gauge choices, it is possible to bring the
metric corrections to the form
\begin{subequations}\label{hone}
\beqa
h_{tt}&=&f_1(r,\theta)\,,\\
h_{tz}&=&f_2(r,\theta)\,,\\
h_{zz}&=&f_3(r,\theta)\gamma_{zz}\,,\\
h_{rr}&=&f_4(r,\theta)\gamma_{rr}\,,\\
h_{\theta\theta}&=&f_5(r,\theta)\gamma_{\theta\theta}\,,\\
h_{\Omega\Omega}&=&f_6(r,\theta)\gamma_{\Omega\Omega}\,,
\eeqa
\end{subequations}
where $\gamma_{\mu\nu}$
is the flat space metric \reef{adapted} to $O(1/R^2)$ and the subindices
${}_{\Omega\Omega}$ denote the coordinates for the $n$-sphere $S^{n}$.
This perturbation involves six functions $f_i(r,\theta)$, but since we are
considering a source with only $T_{tt}$, $T_{tz}$, $T_{zz}$,
actually only three of the functions will be
independent\footnote{Actually, only two if eqs.~\reef{mutaup} are imposed.}.
Finding the complete reduction to three functions is possible but
complicated. Fortunately, for our analysis we will only need to find one
explicit relation among the six functions in \reef{hone}.
The argument is given in appendix~\ref{app:fis} and implies
\beq\label{constr}
f_1-f_3-f_4-f_5-(n-2)f_6=0\,.
\eeq
The off-diagonal perturbation $h_{tz}$,
and hence $f_2$, decouples and can be found separately. Recall also that
the transverse gauge condition $\nabla_\mu \bar h^\mu_\nu=0$ must be
imposed. We will regard it as one more field equation.

The radial dependence of $f_i$ can be fixed by dimensional arguments. If
we separate the zeroth and first order corrections in
$1/R$ we can write
\beq
f_i(r,\theta)=
\frac{r_0^{n}}{r^{n}}\left(f_i^{(0)}(\theta)+\frac{r}{R}f_i^{(1)}(\theta)
\right)\,.
\eeq

To solve for $f_i^{(0)}(\theta)$ note that to zeroth order in $1/R$ we are simply
finding the linearized perturbation created by a {\em straight} linear
distribution of mass, momentum and pressure. Then at this order the $SO(n+2)$
symmetry of the $S^{n+1}$ spheres is unbroken and therefore the
$f_i^{(0)}$ must be constants.
Using \eqref{Tmunu}, these constants are easily found to be
\begin{subequations}
\beqa\label{f0}
f^{(0)}_1&=&\mu+\frac{\tau}{n+1}\,,\\
f^{(0)}_2&=&-p\,,\\
f^{(0)}_3&=&\tau+\frac{\mu}{n+1}\,,\\
f^{(0)}_4&=&f^{(0)}_5=f^{(0)}_6=\frac{\mu-\tau}{n+1}\,.
\label{f04}
\eeqa
\end{subequations}
For later reference, we note that in order to pass to
`Schwarzschild gauge', in which
$h_{\theta\theta}^{(0)}=h_{\Omega\Omega}^{(0)}=0$, we must
change
\beq\label{schgauge}
r\to r-\frac{\mu-\tau}{2(n+1)}\frac{r_0^{n}}{r^{n-1}}\,.
\eeq

Next we solve for $f_i^{(1)}(\theta)$. The $R_{tt}$ equation is
\beq\label{rtt}
{f_1^{(1)}}''+n\cot\theta\, {f_1^{(1)}}'-(n-1)f_1^{(1)}=0\,.
\eeq
We show in appendix~\ref{app:regular} that the only solution of this
equation that is regular on the plane of the ring, \ie at both
$\theta=0,\,\pi$,
is the trivial
one,
\beq\label{f1}
f_1^{(1)}=0\,.
\eeq
The $R_{zz}$ equation for $f_3^{(1)}$ is the same as
\reef{rtt}, so again regularity implies
\beq\label{f3}
f_3^{(1)}=0\,.
\eeq

At this stage, we impose the Einstein equation $R_{r\theta}=0$ which takes the form
\beq
{f_4^{(1)}}'+(n-1){f_6^{(1)}}'+{f_3^{(1)}}'-{f_1^{(1)}}'+(n-
1)\cot\theta({f_6^{(1)}}-{f_5^{(1)}})-\frac{n+2}{n+1}\tau\sin\theta=0\,.
\eeq
If we select the regular solutions $f_1^{(1)}=f_3^{(1)}=0$ and use
\reef{constr}, this equation becomes
\beq\label{rth}
({f_6^{(1)}}-{f_5^{(1)}})'+(n-1)\cot\theta({f_6^{(1)}}-{f_5^{(1)}})-\frac{n+2}{n+1}\tau\sin\theta=0\,.
\eeq
The function $f_6-f_5$ measures the polar
distortion of the $S^{n+1}$-spheres at constant $r$, whose line element is
proportional to
\beq
d\theta^2+(1+f_6-f_5)\sin^2 \theta d\Omega_{n}^2\,.
\eeq
In general we may have $f_6\neq f_5$,
but in order to avoid a conical singularity at the poles of the sphere
the two functions must be equal there,
\beq\label{nocone}
f_6(\theta=0)=f_5(\theta=0)\,,\qquad f_6(\theta=\pi)=f_5(\theta=\pi)\,.
\eeq
In appendix~\ref{app:regular} we show that this condition can only be
satisfied if the inhomogeneous term in the equation \reef{rth} vanishes,
\ie $\tau=0$. So we find that regularity of the solution can be
achieved only if
\beq
T_{zz}=0\,,
\eeq
\ie the tension along the ring must vanish. We have thus reproduced the
result \reef{notzz} as a condition of absence of naked singularities on
the plane of the ring.

Imposing this condition, it is now easy to solve the remaining
equations. Since $T_{tt}$ and $T_{tz}$ do not affect the purely spatial
components of $\bar h_{\mu\nu}$ we must have
$f_4^{(1)}=f_5^{(1)}=f_6^{(1)}$ and then \reef{constr} with \reef{f1},
\reef{f3} imply\footnote{Here, like in \reef{f1}, we are using the fact
that there are no non-trivial and non-singular solutions to the
homogeneous equations with the prescribed boundary conditions.}
\beq
f_4^{(1)}=f_5^{(1)}=f_6^{(1)}=0\,.
\eeq
Finally, the $R_{tz}$ equation
\beq\label{rtz}
{f_2^{(1)}}''+n\cot\theta\, {f_2^{(1)}}'-(n-1)f_2^{(1)}-2np\cos\theta=0\,
\eeq
is easily seen, using arguments like the ones used for \reef{rtt}, to have
\beq
f_2^{(1)}=-p \cos\theta\,
\eeq
as its only regular solution.

If we use the equilibrium value of the boost \reef{eqboost} along
with the relations \eqref{mutaup} and collect the above results,
the solution in the overlap zone $r_0\ll r\ll R$, in transverse gauge, is
\begin{subequations}\label{soltran}
\beqa
g_{tt}&=&-1+\frac{n+1}{n}\frac{r_0^{n}}{r^{n}}\,,
 \label{soltran1}\\
g_{tz}&=&-\frac{\sqrt{n+1}}{n}\frac{r_0^{n}}{r^{n}}
    \left(1+\frac{r\cos\theta}{R}\right)\,,
\label{soltran2}\\
g_{zz}&=&1+\frac{1}{n}\frac{r_0^{n}}{r^{n}}
    \left(1+\frac{2r\cos\theta}{R}\right)+\frac{2r\cos\theta}{R}\,,
\label{soltran3}\\
g_{rr}&=&1+\frac{1}{n}\frac{r_0^{n}}{r^{n}}
    \left(1-\frac{2}{n}\frac{r\cos\theta}{R}\right)-
    \frac{2}{n}\frac{r\cos\theta}{R}\,,
\label{soltran4}\\
g_{ij}&=&\hat g_{ij}\left[
    1+\frac{1}{n}\frac{r_0^{n}}{r^{n}}
    \left(1-\frac{2}{n}\frac{r\cos\theta}{R}\right)-
    \frac{2}{n}\frac{r\cos\theta}{R}
    \right] \,.
\label{soltran5}
\eeqa
\end{subequations}
where in the angular part we have a factor
\beq\label{hatg}
\hat g_{ij}dx^i dx^j= r^2(d\theta^2 +\sin^2\theta d\Omega^2_{n})\,
\eeq
of a round $S^{n+1}$ of radius $r$. The metric, however, has symmetry
$SO(n+1)$ and not $SO(n+2)$ since it depends explicitly on $\theta$.

One can check that the complete linearized solution of the previous
section, \reef{linsol}-\reef{Asol}, reduces to this metric in the
overlap zone. To this effect, one must use the relations \reef{MJA}
between parameters, and the change of coordinates in \reef{coordtrans}.

For the purposes of the following section we pass to
`Schwarzschild gauge' \reef{schgauge},
\beq
r\to r-\frac{1}{2n}\frac{r_0^{n}}{r^{n-1}}\,,
\eeq
in which the solution takes the form
\begin{subequations}\label{solsch}
\beqa
g_{tt}&=&-1+\frac{n+1}{n}\frac{r_0^{n}}{r^{n}}\,,\\
g_{tz}&=&-\frac{\sqrt{n+1}}{n}\frac{r_0^{n}}{r^{n}}
    \left(1+\frac{r\cos\theta}{R}\right)\,,\\
g_{zz}&=&1+\frac{1}{n}\frac{r_0^{n}}{r^{n}}
    \left(1+\frac{r\cos\theta}{R}\right)+\frac{2r\cos\theta}{R}\,,\\
g_{rr}&=&1+\frac{r_0^{n}}{r^{n}}
    \left(1-\frac{2n-1}{n^2}\frac{r\cos\theta}{R}\right)-
    \frac{2}{n}\frac{r\cos\theta}{R}\,,\\
g_{ij}&=&\hat g_{ij}\left(
    1+\frac{1}{n^2}\frac{r_0^{n}}{r^{n-1}R}\cos\theta-
    \frac{2}{n}\frac{r\cos\theta}{R}
    \right) \,.
\eeqa
\end{subequations}

\section{Perturbations of the boosted black string}
\label{sec:nearhor}

\subsection{Setting up the near-horizon perturbation analysis}
\label{sec:pertsetup}

We now turn to the perturbations of the metric near the horizon. These
arise when we force the field of the boosted black string to become like
\reef{solsch} at large distances, \ie we curve the string
into a circle of large but finite radius $R$. In effect, we are placing
the black string in an external potential whose form at large distances
can be read from \reef{solsch}, and which changes the
metric $g^{(0)}_{\mu\nu}$ of the boosted black string by a small
amount\footnote{This $h_{\mu\nu}$ should not be confused with the
one introduced in secs.~\ref{sec:linear} and \ref{sec:equil}.}
\begin{equation}
g_{\mu\nu} = g^{(0)}_{\mu\nu} + h_{\mu\nu}\,
\end{equation}
of order $1/R$.

For the unperturbed metric we take the boosted black string \reef{appab}
with ``critical'' boost parameter \reef{eqboost},
\begin{subequations}
\begin{equation}
g^{(0)}_{tt} = - 1 + \frac{n+1}{n} \frac{r_0^{n}}{r^{n}}
\,,\qquad
g^{(0)}_{tz} = -\frac{\sqrt{n+1}}{n}\frac{r_0^{n}}{r^{n}}
\,,\qquad
g^{(0)}_{zz} = 1 + \frac{1}{n} \frac{r_0^{n}}{r^{n}}
\end{equation}
\begin{equation}
g^{(0)}_{rr} = \left( 1- \frac{r_0^{n}}{r^{n}} \right)^{-1}
\,,\qquad
g^{(0)}_{\theta\theta} = \hat g_{\theta\theta}
\,,\qquad
g^{(0)}_{\Omega\Omega} =
\hat g_{\Omega\Omega}
\end{equation}
\end{subequations}
where $\hat g_{\theta\theta}$ and $\hat g_{\Omega\Omega}$ are the same as given in
\reef{hatg}.

A general discussion of how the perturbations enter in a matched
asymptotic expansion can be found in \cite{Gorbonos:2004uc}, so our discussion will
be very succinct, referring to \cite{Gorbonos:2004uc} for details.
Modes are classified according to their tensorial character upon
coordinate transformations of $S^{n+1}$, and each kind of tensor is then
decomposed into the corresponding spherical harmonics of $S^{n+1}$. The
equations for the modes decouple according to their tensor type and
multipole order. We assume that the symmetry group $SO(n+1)$ of $S^n$ is
unbroken so the only independent components are the scalars $h_{tt}$,
$h_{tz}$, $h_{zz}$, $h_{rr}$, the vector $h_{r\theta}$, and the tensors
$h_{\theta\theta}$, $h_{\Omega\Omega}$.

Given a scalar function, one can form vectors and tensors out of it via
differentiation. For our purposes, it will suffice to consider such
`scalar-derived' vectors and tensors (instead of `pure' vectors and
tensors), which simplifies considerably the analysis---additional pure
tensors, for instance, might be introduced, but they are actually not
needed to solve the problem and since they decouple, they can be
consistently set to zero. This is very convenient, since
Ref.~\cite{Gorbonos:2004uc} shows that for scalar-derived perturbations an
appropriate `no derivative gauge' can be chosen which makes
$h_{r\theta}=0$.

So, up to this point, we must deal with six functions $h_{tt}$,
$h_{tz}$, $h_{zz}$, $h_{rr}$, $h_{\theta\theta}$, $h_{\Omega\Omega}$ of
$r$ and $\theta$. We can still restrict significantly the form of the
perturbations by considering how they arise. The metric
\reef{solsch} at large $r$ can be regarded as providing
the external potential that perturbs the black string. The
only perturbations that are present are proportional to $\cos\theta$,
which is the Legendre polynomial $P_\ell(\cos\theta)$ with $\ell=1$, \ie
they are dipole perturbations. In particular, no monopoles (which would
arise from a spherically symmetric component of the potential) appear.
Since the different multipoles decouple, only dipoles are actually
sourced and the perturbations must all be of the form
\beq
h_{\mu\nu}=\cos\theta\, a_{\mu\nu}(r)\,.
\eeq
Moreover, it can be shown \cite{Gorbonos:2004uc} that, when the tensors with $\ell=1$ are
scalar-derived, we must have
\beq
h_{\theta\theta}=h_{\Omega\Omega}\,.
\eeq
This is enough, then, to reduce the perturbations to the form
\begin{subequations}\label{gmunu}
\begin{equation}
\label{gtta}
g_{tt} = - 1 + \frac{n+1}{n} \frac{r_0^{n}}{r^{n}} +
\frac{\cos \theta}{R} a(r)\,,
\end{equation}
\begin{equation}
g_{tz} = -\frac{\sqrt{n+1}}{n} \left[ \frac{r_0^{n}}{r^{n}} +
\frac{\cos \theta}{R} b(r) \right]\,,
\end{equation}
\begin{equation}
g_{zz} = 1 + \frac{1}{n} \frac{r_0^{n}}{r^{n}} + \frac{\cos \theta}{R}
c(r)\,,
\end{equation}
\begin{equation}\label{grr}
g_{rr} = \left( 1- \frac{r_0^{n}}{r^{n}} \right)^{-1} \left[ 1
+ \frac{\cos \theta}{R} f(r) \right]\,,
\end{equation}
\begin{equation}
\label{gijg}
g_{ij} =  \hat g_{ij}\left[ 1
+ \frac{\cos \theta}{R} g(r) \right]\,,
\end{equation}
\end{subequations}
where $\hat g_{ij}$ is the metric \reef{hatg} of a $S^{n+1}$ of radius $r$.

With this ansatz, the location of the horizon will remain at $r=r_0$ if
the perturbations are finite there. This fixes in part the choice of
radial coordinate, but there still remains some gauge freedom. If we
change
\begin{subequations}\label{gauge}
\beqa
r&\to&r+ \gamma(r)\,\frac{r_0}{R}\,\cos\theta\,,\\
\theta&\to&\theta + \beta(r)\,\frac{r_0}{R}\,\sin\theta\,,
\eeqa
\end{subequations}
with
\beq\label{bgauge}
\beta'(r)=\frac{\gamma(r)}{r^2\left(1- \frac{r_0^{n}}{r^{n}}\right)}
\eeq
then the metric remains in the form above, in particular we still have
$g_{r\theta}=0$ up to order $1/R$. The condition that the horizon stays at
$r=r_0$, \ie\ $r_0\to r_0 +O(1/R^2)$, is
\beq\label{gammao}
\gamma(r_0)=0\,.
\eeq
Under \reef{gauge}, \reef{bgauge}, the perturbation variables change to
\begin{subequations}\label{atogp}
\beqa\label{ag}
a(r)&\to& a(r)-
(n+1)\,\frac{r_0^{n+1}}{r^{n+1}}\,\gamma(r)\,,\\
\label{bg}
b(r)&\to& b(r)-
n\,\frac{r_0^{n+1}}{r^{n+1}}\,\gamma(r)\,,\\
\label{cg}
c(r)&\to& c(r)-
\,\frac{r_0^{n+1}}{r^{n+1}}\,\gamma(r)\,,\\
\label{fg}
f(r)&\to& f(r)+r_0\,
\left(2\gamma'(r)-n\frac{r_0^{n}}{r^{n}}\frac{\gamma(r)}{r\left(1-
\frac{r_0^{n}}{r^{n}}\right)}\right)\,,\\
\label{gpg}
g'(r)&\to& g'(r)+2 \frac{r_0}{r}\,
\left(\gamma'(r)+\frac{r_0^{n}}{r^{n}}\frac{\gamma(r)}{r\left(1-\frac{r_0^{n}}{r^{n}}\right)}\right)
\,.
\eeqa
\end{subequations}
There is furthermore an $r$-independent dipole gauge transformation,
\beq\label{zeromode}
\theta\to\theta + \beta_0\,\frac{r_0}{R}\,\sin\theta\,,
\eeq
with constant $\beta_0$ such that everything remains unchanged except for
\beq
g(r)\to g(r)+2 r_0 \beta_0\,,
\eeq
\ie\ $g$ is redefined by an additive
constant.

We might want to fix the gauge freedom \reef{gauge} by setting one of
the perturbation variables to be a specified function. It is not
advisable to fix neither $a$, $b$ nor $c$ since the condition
\reef{gammao} implies that $a(r_0)$, $b(r_0)$, $c(r_0)$ remain invariant
and we do not know a priori what these values are. A better option is to
fix $f(r)$ or $g(r)$, since their horizon values are not constrained---a
simple-looking gauge choice is $f=g$. However, gauge-fixing
will not be necessary nor useful in our analysis.

It will be more convenient instead to work with variables that are invariant under
\reef{atogp}, such as
\begin{subequations}\label{invar}
\beqa\label{Ainvt}
{\sf A}(r)&=&a(r)-(n+1)\,c(r)\,,\\
{\sf B}(r)&=&b(r)-n \,c(r)\,,\label{Binvt}\\
{\sf F}(r)&=&f(r)+2 r_0\left(\frac{r^{n+1}}{r_0^{n+1}}\;c(r)\right)'
-\frac{n}{\left(1-
\frac{r_0^{n}}{r^{n}}\right)}\;c(r)\,,\label{Finvt}\\
{\sf G}'(r)&=&g'(r)+2
 \frac{r_0}{r}\left(\frac{r^{n+1}}{r_0^{n+1}}\;c(r)\right)'
+\frac{2}{r\left(1- \frac{r_0^{n}}{r^{n}}\right)}\;c(r)\label{Gpinvt}\,.
\eeqa
\end{subequations}
We will refer to these as `gauge-invariant variables', even if some of
the gauge freedom has been already fixed by requiring that
$g_{r\theta}=0$ and by the boundary conditions at $r\to \infty$ and
$r\to r_0$. This way to proceed is similar to the suggestion in
\cite{Kol:2006ga,Kol:2006ux} to postpone gauge-fixing as much as
possible.

\subsection{The master equation and its solution \label{sec:master}}

Our task is to derive a `master equation' for a single gauge-invariant
function, with as few derivatives as possible, so that the rest of the
solution (up to gauge transformations) can be obtained from the solution
to the master field equation. The way we proceed is essentially similar
to the one followed in \cite{Harmark:2003yz,Gorbonos:2004uc}. Ideally the
master equation would be of second order, and in fact
\cite{Harmark:2003yz,Gorbonos:2004uc} managed to reduce
their system of equations involving $h_{tt}$, $h_{rr}$,
$h_{\theta\theta}$, to a single second order ODE for the radial
component of $h_{tt}$. In our case we have two additional functions from
$h_{tz}$ and $h_{zz}$, which we are unable to fully decouple. As a
consequence we have not managed to reduce the master equation to lower
than fourth order \footnote{The reason that it is of fourth instead of
sixth order is ultimately due to leaving the gauge unfixed.}.

While only four equations should suffice for the four gauge-invariant
functions \reef{invar}, in practice it is convenient to work with the six Einstein's
equations corresponding to the vanishing of the Ricci tensor components
$R_{tt}$, $R_{tz}$, $R_{zz}$, $R_{rr}$, $R_{\theta\theta}$, and
$R_{r\theta}$.

The $R_{r\theta}$ equation allows to
eliminate ${\sf F}$ in terms of ${\sf A}$, ${\sf B}$, their first
derivatives, and ${\sf G}'$,
\beqa\label{feq}
{\sf F}&=&\frac{1}{n^3}\Biggl[\frac{n \Big(n (n+1)\, {\sf A}-2 (n+1)\,
{\sf B}\Big)} {\left(1-\frac{r_0^n}{r^n}\right)} + 2 \Big( n^3 r
\,{\sf G}' -2(n+1)(r \,{\sf B}'-\,{\sf B})+n(r\,{\sf A}'-\,{\sf A})
    \Big)\nn\\
&-&\frac{2}{\left(2-\frac{r_0^n}{r^n}\right)}
   \Big( n^3 r \,{\sf G}' -4(n+1)r\,{\sf B}'+n(n+2)r\, {\sf A}'- 2(n-2)(n+1)\,{\sf B}+
   n\left(n^2-2\right) \,{\sf A} \Big)
   \Biggr]
\,. \nn\\
\eeqa
Henceforth this
expression for ${\sf F}$ is plugged into the rest of the equations. Next, using the
$R_{tt}$, $R_{rr}$, $R_{\theta\theta}$ equations we can eliminate ${\sf G}'$
in terms of ${\sf A}$, ${\sf B}$, and their first derivatives,
\beqa\label{gpeq}
{\sf G}'&=&\frac{1}{n^3}\Biggl[\frac{2 n \left[2\, {\sf B}-n (n+2)
\left((n+1)\, {\sf A}-2 \,{\sf B}+r \,{\sf A}'\right)+2 (n+1) r\,
{\sf B}'\right]
   }{(n+1) \,r}\frac{r^n}{r_0^n}\nn\\
&&-\frac{2 (n+1) (n \,{\sf A}-2\, {\sf B})}{r
   \left(1-\frac{r_0^n}{r^n}\right)}
-\frac{(n+2) \left(n\, {\sf A}'-2(n+1)
 \,  {\sf
B}'\right)}{(n+1)}\Biggr]\,.
\eeqa
This can be inserted back into \reef{feq} to similarly find ${\sf F}$ in
terms of only ${\sf A}$, ${\sf B}$, and their first derivatives.

The remaining equations, say, from $R_{tt}$ and $R_{tz}$, reduce now to two
coupled second order ODEs for ${\sf A}(r)$ and ${\sf B}(r)$.
We can easily eliminate one in terms of the other, say ${\sf B}$ in
terms of ${\sf A}$, as
\beqa
\label{Bsol}
{\sf B}&=&\frac{1}{n (n+1)}\frac{r^{n}}{r_0^{n}}
\left(1-\frac{r_0^{n}}{r^{n}}\right)^2 r^3 {\sf A}'''
 + \frac{1}{n (n+1)} \frac{r^{n}}{r_0^{n}}
\left(1-\frac{r_0^{n}}{r^{n}}\right) \left(3
(n+1)-(n+3) \frac{r_0^{n}}{r^{n}}\right)r^2{\sf A}''\nn\\
&&+ \left(2 \frac{r^{n}}{r_0^{n}}-\frac{1}{n}\right)
   \left(1-\frac{r_0^{n}}{r^{n}}\right)r  {\sf A}'
+\left(\frac{3n+2}{n}-2 \frac{r^{n}}{r_0^{n}}\right){\sf A}\,,
\eeqa
and then find a fourth-order ODE for ${\sf A}$, which is
our
master equation,
\begin{eqnarray}
\label{Aeq}
&&\left(1-\frac{r_0^{n}}{r^{n}}\right)^2 r^4 {\sf A}''''(r)
+
\left(1-\frac{r_0^{n}}{r^{n}}\right)
\left[6\left(1-\frac{r_0^{n}}{r^{n}}\right)+4n
\right]
r^3 {\sf A}'''(r) \nn\\ &&
+ \left[5(n+1)^2 - 2(n+3)(n+2)\frac{r_0^{n}}{r^{n}} -
(n^2-7)\frac{r_0^{2n}}{r^{2n}} \right] r^2 {\sf A}''(r) \nn  \\ &&
 + (n^2-1) \left( 2n +1 -
\frac{r_0^{2n}}{r^{2n}} \right) r {\sf A}'(r) -(2n+1)(n^2-1) {\sf A}(r)
 = 0 \,.
\end{eqnarray}
It is useful to note that if, instead, we choose ${\sf B}$ as our master
variable, it satisfies exactly the same fourth-order equation \reef{Aeq}
as ${\sf A}$. Also, ${\sf A}$ is given in terms of ${\sf B}$ by simply
exchanging ${\sf A}\leftrightarrow {\sf B}$ in \reef{Bsol}.

We have found the general solution to \reef{Aeq} through computer-aided
guesswork. It can be expressed in terms of hypergeometric functions, and
a convenient form for the four independent solutions is
\begin{subequations}\label{ui}
\beqa
u_1(r)&=&\ _2F_1\left(-\frac{1}{n},-\frac{n+1}{n};1;
    1-\frac{r_0^n}{r^n}\right)r\,,\\
    \
 u_2(r)&=&\ _2F_1\left(-\frac{1}{n},\frac{n-1}{n};1;
    1-\frac{r_0^n}{r^n}\right)\frac{r_0^n}{r^{n-1}}\,,\\
u_3(r)&=&\ _2F_1\left(\frac{1}{n},\frac{n+1}{n};\frac{n+2}{n};
    \frac{r_0^n}{r^n}\right)\frac{r_0^{n+2}}{r^{n+1}}\,,\\
u_4(r)&=&\ _2F_1\left(\frac{n+1}{n},\frac{2n+1}{n};\frac{3n+2}{n};
    \frac{r_0^n}{r^n}\right)\frac{r_0^{2n+2}}{r^{2n+1}}\,.
\eeqa
\end{subequations}
To check that the four solutions are linearly independent it suffices to
compute their Wronskian in an expansion in powers of $1/r$ and see that
the term at order $r^{-(4n+6)}$ is always non-zero. A general
dimensional argument in \cite{Gorbonos:2004uc} predicts that at leading
order in $1/R$ and multipole order $\ell$ there must appear corrections
$\propto r_0^{n+\ell+1},\, r_0^{2n+\ell+1},\,\dots$. Indeed, terms with
powers
$r_0^{n+2},\,r_0^{2n+2},\,\dots$, consistent with $\ell=1$ dipole
perturbations, appear in the small $r_0$ expansion of
all four terms in \reef{ui}. For $n>2$ these terms are not
analytic in $r_0^n\sim GM/R$.

Since both ${\sf A}$ and ${\sf B}$ satisfy the same equation, they must
be of the form
\begin{subequations}\label{usol}
\beqa
{\sf A}(r)&=& A_1\; u_1(r)  + A_2\; u_2(r)  + A_3\; u_3(r)  +A_4\; u_4(r)
\, ,\\
{\sf B}(r)&=& B_1\; u_1(r)  + B_2\; u_2(r)  + B_3\; u_3(r)  +B_4\; u_4(r)
\, .
\eeqa
\end{subequations}
The constants $B_i$ can in principle be expressed in terms of the $A_i$
using \reef{Bsol}. Then we can also find ${\sf F}$ and ${\sf G}'$ from
\reef{feq} and \reef{gpeq}. However, it is simpler to do this after
fixing the integration constants using the boundary conditions.
Eventually, the solution is fully specified by an appropriate choice of
gauge, \eg\ by fixing $c(r)$. This can not be completely arbitrary,
though, since the boundary conditions at $r\to \infty$ and at the horizon also
fix in part the gauge freedom. So we turn now to imposing boundary
conditions.

\subsection{Boundary conditions}

With our choice of radial coordinate in \reef{grr}, and the partial
gauge-fixing \reef{gammao}, the horizon, if present, remains at $r=r_0$.
We must require that the perturbations do not make this horizon
singular. In particular, the functions $a$, $b$, $c$, $f$, $g$ and
their first derivatives must be finite at $r_0$. Then \reef{Ainvt} and
\reef{Binvt} imply that ${\sf
A}$, ${\sf B}$ and their first derivatives (but {\em not} ${\sf F}$ and ${\sf
G}'$) must be finite at $r_0$. The hypergeometric
functions $_2F_1\left(\alpha,\beta;\gamma;z\right)$ have logarithmic
divergences at $z=1$ whenever $\alpha+\beta=\gamma$. This is the case
for $u_3$ and $u_4$ in
\reef{ui} at $r=r_0$, so finiteness of ${\sf
A}(r_0)$, ${\sf A}'(r_0)$, ${\sf B}(r_0)$ and ${\sf B}'(r_0)$ requires
\beq\label{c3c4}
A_3=A_4=B_3=B_4=0\,.
\eeq
Horizon regularity actually imposes further constraints that will be
dealt with in the next subsection.

On the other hand, the large $r$-asymptotics
of the functions in \eqref{gmunu} are determined
from Eqs.~\reef{solsch}. These imply\footnote{In
principle terms $O\left(r^{-n}\right)$ might be allowed
as well, but they are absent from dipole perturbations.}
\begin{subequations}\label{aasy}
\beqa
a(r)&=&0+O\left(r^{-n-1}\right)\,,\\
b(r)&=&\frac{r_0^n}{r^{n-1}}+O\left(r^{-n-1}\right)\,,\\
c(r)&=&2r+\frac{1}{n}\frac{r_0^n}{r^{n-1}}+O\left(r^{-n-1}\right)\,,\\
f(r) &=&-\frac{2}{n}r+\frac{1}{n^2}\frac{r_0^n}{r^{n-1}}+O\left(r^{-n-1}\right)\,,\\
g(r) &=&-\frac{2}{n}r+\frac{1}{n^2}\frac{r_0^n}{r^{n-1}}+O\left(r^{-n-1}\right)\,,
\eeqa
\end{subequations}
which, using \eqref{Ainvt}, \reef{Binvt}, require that
\begin{subequations}\label{uas}
\beqa
{\sf A}(r)&=&-2(n+1)\, r-\frac{n+1}{n}\frac{r_0^n}{r^{n-1}}+O\left(r^{-n-1}\right)\,,\\
{\sf B} (r) &=&-2n\, r+O\left(r^{-n-1}\right)\,.
\eeqa
\end{subequations}
This is sufficient to determine the
remaining constants $A_1$, $A_2$ and $B_1$, $B_2$ in \reef{usol}.
Expanding the hypergeometric functions for large $r$, we find that
\reef{uas} are satisfied iff
\begin{subequations}\label{c1c2}
\beqa
A_1&=&-\frac{2}{\pi}\frac{(n+1)^2}{n^3}\,\Gamma\left(\frac{1}{n}\right)^2
\Gamma\left(-\frac{n+2}{n}\right)\sin\left(\frac{2\pi}{n}\right)\,,
\label{A1}\\
A_2&=&\frac{3n+4}{n(n+1)}\,A_1\,,\qquad
B_1=\frac{n}{n+1}\,A_1\,,\qquad B_2=\frac{2}{n+1}\,A_1
\eeqa
\end{subequations}
(for $n=2$ we can obtain the correct value $A_1=-9\pi/8$ by analytic
continuation in $n$). The functions ${\sf F}$ and ${\sf G}'$ can now be
integrated explicitly from eqs.~\reef{feq} and \reef{gpeq}. The results
are quoted in appendix~\ref{app:FandGp}.

The boundary conditions \reef{aasy} for $f$ and $g'$ also impose restrictions on the
asymptotic form of the gauge transformations \reef{atogp}, namely
\beq
\gamma'(r)= O\left(r^{-n-1}\right)\,,
\eeq
so
\beq\label{asygam}
\gamma(r)= k\, r_0+O\left(r^{-n}\right)\,,
\eeq
with dimensionless constant $k$. With this $\gamma$
in \reef{cg}, $c$ changes as
\beq\label{asygc}
 c \to c - k \frac{r_0^{n+2}}{r^{n+1}}+O\left(r^{-2n-1}\right)\,,
\eeq
which does not affect the
term $\propto r^{-2n+1}$ in the expansion of $c(r)$ for large $r$. So
this term must be fixed once the
gauge-invariant functions are determined. 
Using eqs.\ (6.12c), (6.12d), (6.20c)-(6.20e) and the solution 
for ${\sf F}$ and ${\sf G}'$ at large $r$ ($e.g.$, expand eqs.\ (D.2) and (D.3)),
it is easy to verify that the large-$r$ expansion of $c(r)$
is in fact
\beq\label{casy}
c(r)=
2r+\frac{1}{n}\frac{r_0^n}{r^{n-1}}+k_1\frac{r_0^{n+2}}{r^{n+1}}-
\frac{n^2+2n-1}{2n^2(n-2)}\frac{r_
0^{2n}}{r^{2n-1}}+O\left(r^{-2n-1}\right)\,.
\eeq
So one term is fixed in $c(r)$ beyond those
given in \reef{aasy}, but the constant $k_1$, and all
$O\left(r^{-2n-1}\right)$ terms, do change in general under \reef{asygc}.
For $n=2$ the term $\propto r_0^{2n}$ is replaced by
$-\frac{3}{4}\frac{r_0^4}{r^3}\log r$.

In addition, the asymptotic behavior of $g(r)$ removes some more gauge
freedom. It requires that $g(r)+2r/n\to 0$ as $r\to\infty$, so if we fix
completely the gauge symmetry \reef{gauge}, then the freedom to add a
constant to $g(r)$ using \reef{zeromode} disappears.

\subsection{Horizon regularity}

With our choice of \reef{c3c4} the perturbation variables remain finite
at the horizon $r=r_0$.
Horizon regularity also requires that the surface gravity and the
angular velocity be well-defined and uniform over the horizon
\cite{Racz:1992bp}.
To study whether it is possible to
meet these conditions, we first note that the
horizon is generated by the null orbits of the Killing vector field
\beq
\chi=\frac{\partial}{\partial t}+\Omega_H \frac{\partial}{\partial \psi}
=\frac{\partial}{\partial t}+\Omega_H R\frac{\partial}{\partial z}\,,
\eeq
where the angular velocity is
\beqa
\label{OmH}
\Omega_H&=&-\frac{1}{R}\left(\frac{g_{tz}}{g_{zz}}\right)_{r_0}\nn \\
&=&\frac{1}{\sqrt{n+1}}\frac{1}{R}\left(1+\frac{\cos\theta}{R}\left(b(r_0)-
\frac{n}{n+1}c(r_0)\right)\right)\,.
\eeqa
If $\Omega_H$ is to be uniform over the horizon we must have
\beq\label{bo}
c(r_0)=\frac{n+1}{n}b(r_0)\,.
\eeq
Furthermore, if $\chi$ is null at $r=r_0$ then it easily follows that
$(g_{tt}+\Omega_H g_{tz})_{r_0}=0$. Together with the
$\theta$-independence of $\Omega_H$, this requires
\beq\label{bo2}
b(r_0)=n a(r_0)\,.
\eeq
Our solution of the equations, \reef{usol}, \reef{c3c4}, \reef{c1c2},
determines ${\sf A}$
and ${\sf B}$ at the horizon,
\begin{subequations}
\beqa
{\sf A}(r_0)&=&A_1\;r_0\;\frac{(n+2)^2}{n(n+1)}=a(r_0)-(n+1)\;c(r_0)\,,\\
 {\sf B}(r_0)&=&A_1\;r_0\;\frac{n+2}{n+1}=b(r_0)-n\; c(r_0)\,,
\eeqa
\end{subequations}
where, to avoid clutter, we do not substitute the explicit value
\reef{A1} of $A_1$. These two equations together with \reef{bo}
completely determine the horizon values
\beqa\label{ahor}
a(r_0)&=&-A_1\;r_0\;\frac{n+2}{n^2(n+1)}\,,\nn\\
b(r_0)&=&-A_1\;r_0\;\frac{n+2}{n(n+1)}\,,\\
c(r_0)&=&-A_1\;r_0\;\frac{n+2}{n^2}\,.\nn
\eeqa
Happily, \reef{bo2} is also satisfied with these values.

Next, the surface gravity $\kappa$, defined by
\beq
\label{kap}
\kappa \chi^a=\chi^b\nabla_b \chi^a\,,
\eeq
is
\beq
\label{kapa}
\kappa=\frac{n}{2}\sqrt{\frac{n}{n+1}}\frac{1}{r_0} \left[
1-\frac{r_0
(n+1)}{2n^2}\frac{\cos\theta}{R}
\left(a'(r_0)-\frac{2b'(r_0)}{n}+\frac{c'(r_0)}{n+1}+\frac{n^2
f(r_0)}{(n+1)r_0}\right)\right]
\eeq
which is uniform over the horizon only if
\beq\label{fo}
a'(r_0)-\frac{2b'(r_0)}{n}+\frac{c'(r_0)}{n+1}+\frac{n^2
f(r_0)}{(n+1)r_0}=0
\,.
\eeq
In contrast to $a(r_0)$, $b(r_0)$ and $c(r_0)$, neither of the
quantities entering this equation (nor $g(r_0)$ and $g'(r_0)$) are
invariant under \reef{gauge} and hence are not
determined until we fix the gauge. However, it is easy to check
that when \reef{gammao} holds, equation \reef{fo} is invariant under
\reef{gauge} and so it must be satisfied independently
of the gauge choice. Although ${\sf F}$ and ${\sf G}'$ are in general
singular at $r_0$, using \eqref{ahor} in their solutions \reef{Fsol} and
\reef{Gpsol} we can readily
check that $f$ and $g'$ remain finite there. The value of $f$ at the
horizon that results,
\beq
f(r_0)=-r_0 c'(r_0)+A_1 r_0\frac{n^2+n-2}{n^3}\,\label{fhor}
\eeq
is such that using eqs.\ (6.12a), (6.12b) and the solution 
for ${\sf A}$ and ${\sf B}$, eq.\ (6.35) is identically satisfied and so the surface
gravity is constant on the horizon. The actual values of
$g(r_0)$ and $g'(r_0)$ are not needed for checking the regularity of the horizon.

\subsection{The complete solution}

At this stage, we have given the complete explicit form for the
functions ${\sf A}$, ${\sf B}$, ${\sf F}$, ${\sf G'}$ in
eqs.~\reef{usol}, \reef{c3c4}, \reef{c1c2}, \reef{Fsol}, \reef{Gpsol}.
Finding ${\sf G}$ requires an additional integration that we have not
been able to perform explicitly in closed analytic form. But up to this
quadrature, the solution is fully specified except for a choice of gauge
that fixes the freedom in \reef{gauge}. From \reef{invar}
we see that this amounts to specifying the function $c(r)$, which is
only constrained to preserve the asymptotic form in \reef{casy} and the
horizon value determined in \reef{ahor}. As explained above, once this
is performed the boundary conditions imply that the symmetry \reef{zeromode}
is also fixed.

A simple-looking way of fixing the gauge is to set $f=g$. In this case,
since we have the explicit solution for ${\sf F}$ and ${\sf G'}$, the
function $c(r)$ is determined from the compatibility of equations
\reef{Finvt} and \reef{Gpinvt}. These give a second order ODE for
$c(r)$, which, except for $n=1$ (see appendix \ref{app:5D}), is
complicated to solve. One should show that it is possible to find
solutions to this equation that satisfy the boundary conditions for $c$
at large $r$ while preserving regularity of the horizon, which may not
be easy to prove. In fact this is the typical situation when one chooses
the gauge {\em a priori}, without knowing whether it is actually
compatible with regularity requirements.

In our approach, instead, leaving the gauge symmetry \reef{gauge}
unfixed until the end has allowed us to bypass this complication. We
have managed to show that any choice of $c(r)$ with the appropriate
boundary behavior produces a complete solution that is regular on and
outside the horizon. For instance, we may truncate $c(r)$ to the terms
shown explicitly in \reef{casy}, and choose $k_1$ so that \reef{ahor} is
satisfied. Quite possibly, other choices may be simpler or more natural,
so, not needing to, we shall not dwell more on this issue.

The form of our complete solution and the drastic simplifications that
occur in $n=1$ (see appendix \ref{app:5D}) but not in any other $n>1$
suggest that, barring the discovery of a miraculous coordinate system,
a closed exact analytic solution for a black ring exists only in five
dimensions. Nevertheless, we have managed to provide a very
explicit form of the approximate solution for a thin black ring in any
dimension $D\geq 5$.

\subsection{Properties of the corrected solution}

From our results we can readily see that the area, surface gravity and angular
velocity receive no modifications in $1/R$. The reason is simple: the
perturbations are only of dipole type, with no monopole terms. But a
dipole can not change the total area of the horizon, only its shape.
This is true both of the shape of the $S^{n+1}$ as well as the length of
the $S^1$, which can vary with $\theta$ but on average (\ie when
integrated over the horizon) remains constant. So ${\cal A}$ is not
corrected. The surface gravity and angular velocity can not be corrected
either. They must remain uniform on a regular horizon, so, since the
dipole terms vanish at $\theta=\pi/2$, no corrections to $\kappa$
and $\Omega_H$ are possible.
Following the arguments advanced in sec.~\ref{sec:matchasexp}, we can use the 1st law,
\beq
\label{firstlawa}
dM=\frac{\kappa}{8\pi G}d\mathcal{A}+\Omega_H dJ
\eeq
and the Smarr formula
\beq
\label{smarr}
(n+1)M=(n+2) \left( \frac{\kappa {\cal A}}{8\pi G}+\Omega_H J \right)
\eeq
to conclude that if $\mathcal{A}$, $\kappa$ and $\Omega_H$ are not
corrected, then $M$ and $J$ cannot be corrected either\footnote{In
five-dimensions ($n=1$) there {\em are} corrections to this order. Their
origin is discussed in appendix~\ref{app:5D}.}. So the function
$\mathcal{A}(M,J)$ obtained in \reef{MJA} is indeed valid including the
first order in $1/R$. It is interesting to observe that this conclusion
could be drawn already when the asymptotic form of the metric in the
overlap zone, \reef{adapted}, is seen to include only
dipole terms at order $1/R$.

The length of the $S^1$ along $\psi$ is corrected by $c(r)$ and since
$c(r_0)\neq 0$, this length does vary at different latitudes $\theta$ on
the horizon. Eqs.~\reef{c1c2} and \reef{ahor} imply that $c(r_0)>0$, so
the $S^1$ circle is longer along the outer rim of the ring ($\theta=0$:
the pole looking towards infinity) than along its inner rim
($\theta=\pi$: looking towards the inner disk), as expected. Amusingly,
the distortion of the spheres $S^{n+1}$ at constant $\psi$ on the horizon
is measured by $g(r_0)$ and therefore is a gauge-dependent property,
which is only determined once a choice for $c(r)$ is made. With an
appropriate gauge choice, $g(r_0)$ can take any prescribed value---
positive, negative or zero. In the latter case the $S^{n+1}$ at the horizon remain
metrically round, although $g(r)$ is negative at large $r$.
We have illustrated this point in
appendix~\ref{app:5D}.

\section{Higher-dimensional black rings vs MP black
holes}\label{sec:phases}

We now turn to study the thermodynamics of the thin black ring in the ultraspinning
regime and to compare it to the exact known results for MP black holes.

\subsubsection*{Black ring}
For the convenience of the reader we collect here the entire thermodynamics from
\reef{MJA}, \eqref{OmH}, \eqref{kapa}:
\begin{subequations}
\label{tbrthermo}
\begin{equation}
\label{tbrthermo1}
M=\frac{\Omega_{n+1}}{8 G}\,R\, r_0^{n}(n+2) \spa
S=\frac{\pi\,\Omega_{n+1}}{2G}  R\,r_0^{n+1}
\sqrt{\frac{n+1}{n}} \spa T_H = \frac{n}{4\pi} \sqrt{
\frac{n}{n+1}} \frac{1}{r_0}\,,
\end{equation}
\begin{equation}
\label{tbrthermo2}
J=\frac{\Omega_{n+1}}{8 G}\,R^2\, r_0^{n}\sqrt{n+1}\,,\qquad
\Omega_H = \frac{1}{\sqrt{n+1}} \frac{1}{R}\,.
\end{equation}
\end{subequations}
As we have seen in the previous section,
these results are valid up to $O(r_0^2/R^2)$ corrections.

\subsubsection*{MP black hole}
For the MP black hole, exact results can be obtained for all values of
the rotation.
The two independent parameters specifying the solution are the mass
parameter $\mu$ and the rotation parameter $a$, from which the horizon
radius $r_0$ is found as the largest (real) root of the
equation
\beq\label{mueq}
\mu = (r_0^2 + a^2) r_0^{n-1}\,.
\eeq
In terms of these parameters the thermodynamics take
the form \cite{Myers:1986un}
\begin{subequations}\label{TMPJOm}
\begin{equation}
\label{TMP}
 M = \frac{ (n+2)  \Omega_{n+2} \, \mu}{16 \pi G}
\,,\qquad  S = \frac{\Omega_{n+2} \, r_0 \, \mu}{4 G} \,,\qquad
T_H = \frac{1}{4 \pi} \left( \frac{2 r_0^n}{\mu} + \frac{n-1}{r_0}
\right)\,,
\end{equation}
\begin{equation}\label{JOm}
J = \frac{ \Omega_{n+2}\, a \, \mu }{ 8 \pi G}
\,,\qquad \Omega_H = \frac{a \, r_0^{n-1}}{\mu}\,.
\end{equation}
\end{subequations}
Observe that $a=\frac{n+2}{2} \frac{J}{M}$ bears similarity with \reef{RJM}.
In fact, an important simplification occurs in the ultra-spinning regime
of $J\to\infty$ with fixed $M$, which corresponds to
$a \rightarrow \infty$. Then \reef{mueq} becomes
\begin{equation}
\label{uslimit}
\mu \rightarrow a^2 r_0^{n-1}
\end{equation}
leading to simple expressions for the eqs.~\reef{TMPJOm} in terms of $r_0$ and
$a$, which in this regime play roles analogous to those of $r_0$ and $R$ for the
black ring. Specifically, $a$ is a measure of the size of the horizon along the
rotation plane and $r_0$ a measure of the size transverse
to this plane \cite{Emparan:2003sy}. In fact, in this limit
\beq
\label{TMP2}
 M \to \frac{ (n+2)  \Omega_{n+2}}{16 \pi G}\; a^2 r_0^{n-1}
\,,\qquad S \to \frac{\Omega_{n+2}}{4 G}\;a^2 r_0^{n} \,,\qquad
T_H \to \frac{n-1}{4 \pi r_0}
\end{equation}
take the same form as the expressions characterizing a black membrane extended
along an area $\sim a^2$ with horizon radius $r_0$.
This identification\footnote{The entropy corresponds precisely to a membrane of
planar area $\frac{\Omega_{n+2}}{\Omega_n}a^2$. This value also
gives the precise membrane mass once the dimension dependence of the
mass normalization is taken into account.} lies at the core of
the ideas in \cite{Emparan:2003sy}, and we shall use it extensively in the next section.
The reader may rightly wonder what happens to
\begin{equation}\label{JOm2}
J \to \frac{ \Omega_{n+2}}{ 8 \pi G}\;a^3 r_0^{n-1}
\,,\qquad \Omega_H \to \frac{1 }{a}\,,
\end{equation}
under this identification. Both turn out to disappear, since the
black membrane limit is approached in the region near the axis of rotation of
the horizon and so in the limit the membrane is static.
Also observe that since eq.~\reef{mueq} is quadratic in $a$, the value
\reef{uslimit}, and hence also \reef{TMP2} and \reef{JOm2}, are again valid
up to $O(r_0^2/a^2)$ corrections.

The transition to this membrane-like regime is signaled by a qualitative
change in the thermodynamics of the MP
black holes.
At
\beq\label{onset}
\left(\frac{a}{r_0}\right)_{\rm mem}=\sqrt{\frac{n+1}{n-1}}\,,
\eeq
the temperature reaches a minimum and
$\left(\partial^2 S/\partial J^2\right)_M$
changes sign\footnote{We do not believe that
this sign change is associated to any dynamical instability. Rather, as
discussed below, we expect the
GL-like instability to happen at a larger value of
$a/r_0$.}. For $a/r_0$ smaller than this value, the thermodynamic quantities
of the MP black holes such as $T$ and $S$ behave similarly to those of the Kerr
solution and so we should not expect any membrane-like behavior.
However, past this point they rapidly approach the membrane results.

\subsubsection*{Comparison}

A meaningful comparison between dimensionful magnitudes of two solutions
requires the introduction of a common scale. Then the comparison is made
between dimensionless quantities obtained by factoring out
this scale. Since classical General Relativity does not possess any
intrinsic scale, we must take this to be one of the physical parameters
of the solutions, which we conveniently take to be the mass. Thus
we introduce dimensionless quantities for the spin $j$, the area $a_H$, the angular
velocity $\omega_H$ and the temperature $\mathfrak{t}_H$ as
\begin{subequations}\label{jaot}
\beq
\label{jaHdef}
j^{n+1}=c_j\,
\frac{J^{n+1}}{GM^{n+2}} \,,\qquad
a_H^{n+1}=c_a\,\frac{\mathcal{A}^{n+1}}{(GM)^{n+2}}
~,
\eeq
\beq
\label{otdef}
\omega_H =c_\omega \, \Omega_H
(GM)^{\frac{1}{n+1}} \,,\qquad
\mathfrak{t}_H = c_{\mathfrak{t}}\, (GM)^{\frac{1}{n+1}}\, T_H\,,
\eeq
\end{subequations}
where the numerical constants are
\begin{subequations}
\beq
c_j =\frac{\Omega_{n+1}}{2^{n+5}}\frac{(n+2)^{n+2}}{(n+1)^{\frac{n+1}{2}}}\,,\qquad
c_a=\frac{\Omega_{n+1}}{2(16\pi)^{n+1}}(n+2)^{n+2}
\left(\frac{n}{n+1}\right)^{\frac{n+1}{2}}\,,
\eeq
\beq
c_\omega=\sqrt{n+1}\left(\frac{n+2}{16}\Omega_{n+1}\right)^{-\frac{1}{n+1}}
\,,\qquad
c_{\mathfrak{t}} =  4\pi \sqrt{ \frac{n+1}{n}}
\left( \frac{n+2}{2} \Omega_{n+1}\right)^{-\frac{1}{n+1}}
\,.
\eeq
\end{subequations}
These are defined so that for $D=5$ ($n=1$) the
conventional choice is reproduced and so that some of the formulae are simplified.
Studying the entropy, or the area $\mathcal{A}$, as a function
of $J$ for fixed mass is equivalent to finding the function $a_H (j)$.
Similarly, we are interested in the quantities $\omega_H$ and
$\mathfrak{t}_H$ as functions of $j$, which we take as our control parameter.

The quantities \eqref{jaot} can depend only on
the dimensionless ratios $r_0/R$ and $r_0/a$
specifying the solutions. To translate this dependence in terms of the variable
$j$, we can use for thin black rings and MP black holes respectively,
the relations
\begin{subequations}
\beqa
 j ^{n+1} &=&
2^{-(n+2)}\left(\frac{R}{r_0}\right)^{n}\,,\\
j ^{n+1} &=& \frac{\pi}{(n+1)^\frac{n+1}{2}}
\frac{\Omega_{n+1}}{\Omega_{n+2}}
\frac{\left(\frac{a}{r_0}\right)^{n+1}}{1+\left(\frac{a}{r_0}\right)^2}
\stackrel{a\to \infty}{\longrightarrow}
\frac{\pi}{(n+1)^\frac{n+1}{2}}
\frac{\Omega_{n+1}}{\Omega_{n+2}}\left(\frac{a}{r_0}\right)^{n-1}\,.
\eeqa
\end{subequations}
These expressions make it clear that in the regime $j\gg 1$ the black
rings become thin and the MP black holes pancake along the rotation
plane. According to \reef{onset}, the onset of this behavior for the
latter happens at
\beq\label{jmem}
j_{\rm mem}^{^{n+1}} =
\frac{\pi}{2n (n-1)^\frac{n-1}{2}}
\frac{\Omega_{n+1}}{\Omega_{n+2}}\,.
\eeq

We can now obtain the asymptotic phase curves
for the thin black ring,
\beq\label{ajring}
a_H =  2^{\frac{n-2}{n(n+1)}}\frac{1}{j^{1/n}}\,, \qquad
\omega_H = \frac{1}{2j} \,,\qquad
\mathfrak{t}_H = 2^{\frac{2-n}{n(n+1)}} n \, j^{1/n}
\,.
\eeq
Similarly, for the MP black hole in the ultra-spinning regime
\begin{subequations}
\beq\label{mpaj}
a_H\to 2^{\frac{3}{n+1}}\left(\frac{\pi\Omega_{n+1}}{\Omega_{n+2}}\right)^{\frac{1}{n-1}}
\frac{n^{1/2}}{(n+1)^\frac{n+1}{2(n-1)}}\frac{1}{j^{2/(n-1)}}
~ ,
\eeq
\beq
\omega_H \to \frac{1}{j}
~ , \qquad
\mathfrak{t}_H \to 2^{-\frac{3}{n+1}} \frac{(n-1)(n+1)^{\frac{n+1}{2(n-1)}}}{n^{1/2}}
\left(\frac{\Omega_{n+2}}{\pi \Omega_{n+1}}\right)^{\frac{1}{n-1}} j^{2/(n-1)}
\, .
\eeq
\end{subequations}

In the following we denote
the results for the thin black ring with $^{(r)}$ and for the ultraspinning MP black
holes with $^{(h)}$, and generally omit numerical prefactors.

Starting with the reduced area function we see that
\beq
\label{aHrh}
a_H^{(r)}\sim \frac{1}{{j}^{1/n}}\,,\qquad a_H^{(h)}\sim \frac{1}{{j}^{2/(n-1)}}\,,
\eeq
and so, for any $D=4+n\geq 6$, the area decreases faster for
MP black holes than for black rings: {\it black rings dominate
entropically in the ultraspinning regime}.  Note also that in both
cases as the dimension grows, the fall-off with $j$ becomes slower, asymptoting
to $a_H \to j^0$ when $n\to \infty$ (this includes the numerical prefactor).

\begin{figure}[ht!]
\begin{picture}(0,0)(0,0)
\put(417,16){$j$}
\put(38,252){$a_H$}
\put(65,3){$j_{\rm mem}$}
\end{picture}
\centerline{\includegraphics[width=14cm]{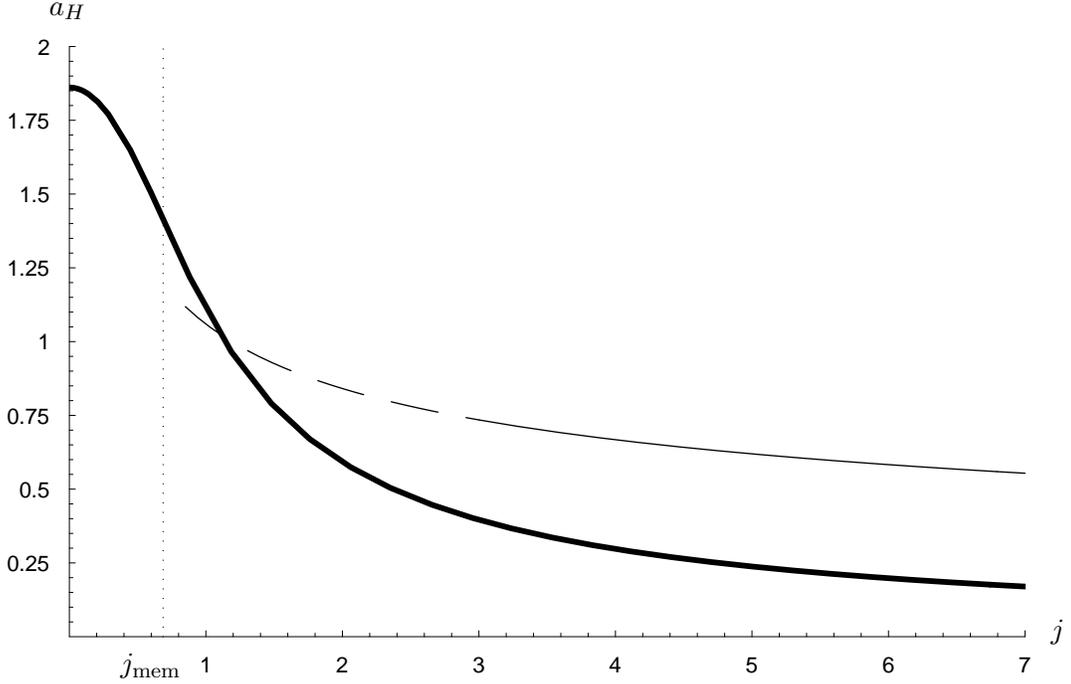}}
\caption{\small Area vs spin for fixed mass, $a_H(j)$, in seven
dimensions. The thin curve is our result for thin black rings, valid at
large $j$ and extrapolated (dashed) down to $j\sim O(1)$. The thick
curve is the exact result for the MP black hole. The vertical dotted
line intersects this curve at the inflection point $j=j_{\rm
mem}=2^{1/4}/\sqrt{3}$, $a_H=\sqrt{2}$. It signals the onset at larger
$j$ of membrane-like behavior for MP black holes. Eq.~\reef{7D} gives
the asymptotic form of the curves. The same qualitative features appear
for all $D\geq 6$.}
\label{fig:MPBR}
\end{figure}

For illustration, we plot in Fig.~\ref{fig:MPBR} the curves in $D=7$
($n=3$), whose asymptotic form, including
numerical factors, is
\beq\label{7D}
a_H^{(r)}\to \frac{2^{1/12}}{j^{1/3}}\,,\qquad
a_H^{(h)}\to \frac{2^{1/4}}{j}\,.
\eeq

The ratio $\omega^{(h)}_H/\omega^{(r)}_H=2$, which holds for all $D\geq
6$, is reminiscent of the factor of 2 in Newtonian mechanics between the
moment of inertia of a wheel (\ie a ring) and a disk (\ie a pancake) of
the same mass and radius, which implies that the disk must rotate twice
as fast as the wheel in order to have the same angular momentum. Irrespective of
whether this is an exact analogy or not, the fact that $\omega^{(r)}_H<\omega^{(h)}_H$
is clearly expected from this sort of picture.

For the temperatures we find
\beq \label{tempsa}
\mathfrak{t}^{(r)}_H \sim j^{1/n} \spa \mathfrak{t}^{(h)}_H \sim
j^{2/(n-1)}\, , \eeq
so the thin black ring is colder than the MP
black hole. In fact, the picture suggested above leads to the
following argument: if we put a given mass in the shape of a wheel
of given radius, then we get a thicker object than if we put it in
the shape of a pancake of the same radius. More precisely, the
spread on the rotation plane is in both cases $\sim j$, but the
thickness $r_0$ is, for fixed mass, \beq r_0^{(r)}\sim
j^{-1/n}\,,\qquad r_0^{(h)}\sim j^{-2/(n-1)}\,. \eeq Then, the
expressions in \eqref{tempsa} follow immediately from the fact that
the temperature is inversely proportional to the thickness $T_H\sim
1/r_0$. Moreover, observing that $T_H {\cal A}$ is independent of
$j$ for fixed mass, the converse of this argument `explains' why the
black ring has higher entropy than the MP black hole of the same
mass and spin.

\section{Towards a complete phase diagram}
\label{sec:complete}

The curve $a_H(j)$ at values of $j$ outside the domain of validity of our computation
correspond to the regime where the gravitational self-attraction of the
ring is important. At present, no analytical methods are known that can
deal with such values $j\sim O(1)$. The precise form of the curve in this
regime may require numerical solutions.

However, it is possible to complete this curve and other features of the
phase diagram, at least qualitatively, by combining a number of
observations and reasonable conjectures about the behavior of MP black
holes at large rotation and using as input the presently known phase
structure of Kaluza-Klein black holes.

\subsection{GL instability of ultra-spinning MP black hole}

In the ultraspinning regime in $D\geq 6$, MP black holes
approach the geometry of a black membrane $\approx \bbr{2} \times
S^{D-4}$ spread out along the plane of rotation \cite{Emparan:2003sy}.
In the previous section we have already
observed that the extent of the black hole along the plane is approximately
given by the rotation parameter $a$, while the `thickness' of the
membrane, \ie the size of its $S^{D-4}$, is given by the parameter $r_0$.
For $a/r_0$ larger than the critical value \reef{onset}
we expect that the dynamics of these black holes
is well-approximated by a black membrane compactified on a square torus
$\T^2$ with side length $L\sim a$ and with $S^{D-4}$ size $\sim r_0$.
The angular velocity of the black hole is always moderate, so it will
not introduce large quantitative
differences, but note that the rotational axial symmetry of the MP black
holes translates into only one translational symmetry along the $\T^2$,
the other one being broken.

\begin{figure}[ht!]
\centerline{\includegraphics[width=12cm]{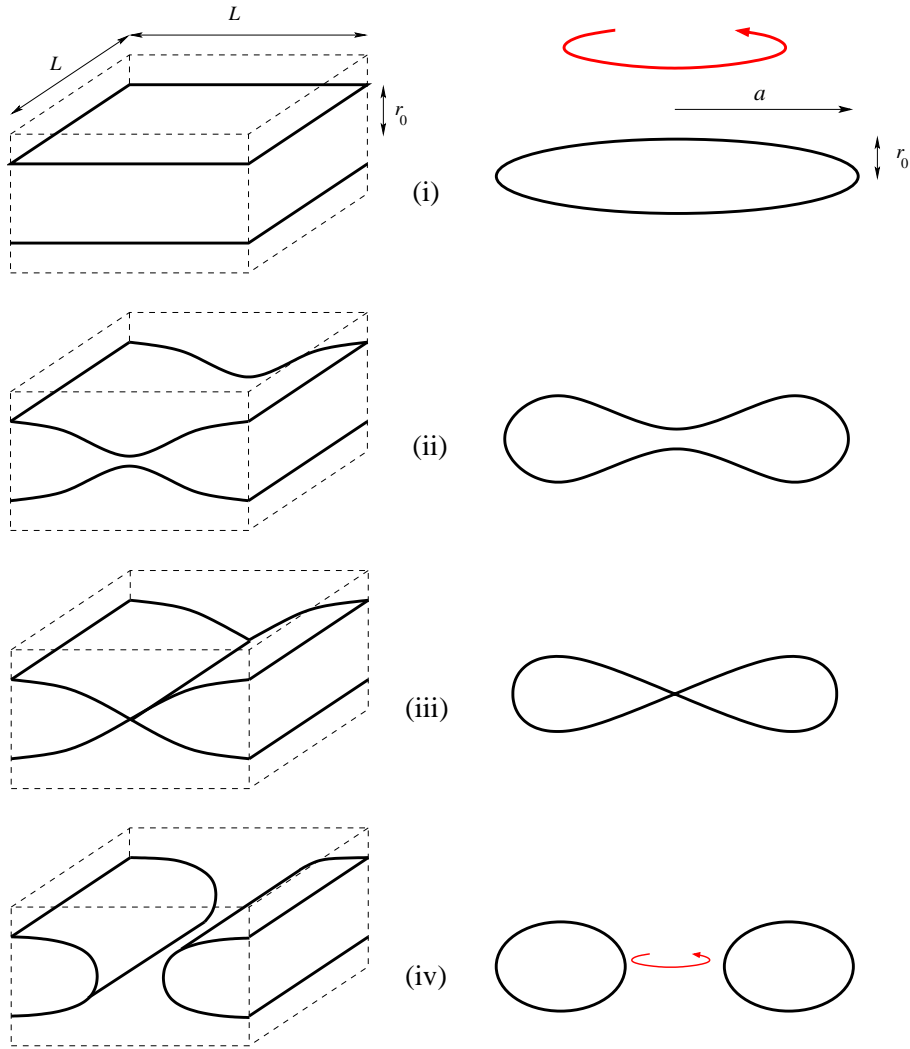}}
\caption{\small Correspondence between phases of black membranes wrapped
on a $\T^2$ of side $L$ (left) and fastly-rotating MP black holes with rotation
parameter $a\sim L\geq r_0$ (right: must be rotated along a vertical axis):
(i) Uniform black membrane and MP black hole.
(ii) Non-uniform black membrane and pinched black hole.
(iii) Pinched-off membrane and black hole. (iv) Localized black string
and black ring.}
\label{fig:membranes}
\end{figure}

Using this analog mapping of membranes and fastly rotating MP black
holes, Ref.~\cite{Emparan:2003sy} argued that the latter should exhibit
a Gregory-Laflamme-type of instability. Furthermore, it is known that
the threshold mode of the GL instability gives rise to a new branch of
static non-uniform black strings and branes
\cite{Gregory:1988nb,Gubser:2001ac,Wiseman:2002zc}. In correspondence
with this, Ref.~\cite{Emparan:2003sy} argued that it is natural to
conjecture the existence of new branches of axisymmetric `lumpy' (or
`pinched') black holes, branching off from the MP solutions along the
stationary axisymmetric zero-mode perturbation of the GL-like
instability,

We intend to develop further this analogy, and draw a correspondence between the phases
of black membranes and the phases of higher-dimensional black holes, as illustrated
in Fig.~\ref{fig:membranes}. Observe that we only consider the inhomogeneities of the
membrane along one of the brane directions, since the other ones do not have a
counterpart for rotating black holes: they would break axial symmetry
and hence would be radiated away. Other limitations of the analogy will
be discussed in sec.~\ref{subsec:conject}.

\subsection{Phase diagram of black membranes and strings on $\CM^{D-2} \times \T^2$}

Having this correspondence between the phases of the two systems, we can
try to import, at least qualitatively, the known phase diagram of black
membranes on $\CM^{D-2} \times \T^2$ onto the phase diagram of rotating
black objects in $\CM^D$. This requires that, first, we establish the
map between quantities on each side of this correspondence. Second, we
must collect the available information about the former in an
appropriate form.

Mapping the results for Kaluza-Klein black holes on $\CM^{D-2}
\times \T^2$ to rotating black objects in $\CM^D$ requires that in
both cases we choose to fix the overall scale in the same manner.
For Kaluza-Klein black holes the preferred scale is usually the
circumference $L$ of the asymptotic circles. For rotating black
holes, instead, we have chosen the mass as the common scale to
define dimensionless magnitudes. We therefore introduce for
Kaluza-Klein black holes on $\T^2$ the dimensionless `length' $\ell$
and dimensionless `horizon area' $a_H$
\begin{equation}
\label{eladef}
\ell^{n+1} = \frac{L^{n+1}}{16 \pi G M} \,,\qquad a_H^{n+1} = \frac{1}{4^{n+3} 16 \pi}
\frac{{\cal A}^{n+1}}{(GM)^{n+2}}\,,
\end{equation}
where $L$ is the side length of the square torus $\T^2$.

For unit mass, the quantities $\ell$ and $j$ measure the (linear) size
of the horizon along the torus or rotation plane, respectively. Then
$a_H(\ell)$ for KK black holes on $\CM^{n+2} \times \T^2$ is analogous
(up to constants) to $a_H(j)$ for rotating black holes in $\CM^{n+4}$.
More precisely, although the normalization of magnitudes in
\reef{jaHdef} and \reef{eladef} are different, the functional dependence
of $a_H$ on $\ell$ or $j$ must be parametrically the same in both
functions, at least in the regime where the analogy is precise.

What is then known about the function $a_H(\ell)$ for the different KK phases?
Most of the information about black holes in KK spacetimes has been
obtained for solutions on a KK circle instead of on $\T^2$ (see e.g. the reviews
\cite{Kol:2004ww,Harmark:2007md}). However, this is enough for our purposes since we
are only considering the possibility of non-uniformity along one of the torus
directions. Then, phases of black membranes/localized black strings
on $\CM^{n+2} \times \T^2$
are simply obtained by adding a flat direction to the phases of black strings/localized black
holes on $\CM^{n+2} \times S^1$. In appendix \ref{app:torusphases} we briefly review
what is known about these phases and provide a translation dictionary to
obtain the corresponding results for phases on $\T^2$ that we use below.

For the uniform black membrane (ubm) in $ 4 +n$ dimensions the function $a_H(\ell)$
has the behavior
\begin{equation}
\label{aHmem}
a_H^{\rm ubm} (\ell) \sim \ell^{- \frac{2}{n-1}}\,.
\end{equation}
This exhibits exactly the same functional form \reef{aHrh} as $a_H(j)$
for the MP black hole in the ultra-spinning limit. Furthermore, for the localized
black string (lbs) in $ 4 +n$ dimensions one finds
\begin{equation}
\label{aHstr}
a_H^{\rm lbs} (\ell) \sim \ell^{- \frac{1}{n}}   \qquad (\ell
\rightarrow \infty)\,,
\end{equation}
which shows again the same functional form \eqref{aHrh} as $a_H (j)$ of the
black ring in the large $j$ limit (the fact that we have considered a
static, instead of boosted, localized black string only affects
numerical factors).
These results illustrate the quantitative aspects of the analogy in the large $
j$, or large $\ell$, regime.

The most important application of the analogy, though, is to non-uniform
membrane phases, providing information about the phases of pinched
rotating black holes and how they connect to MP black holes and black
rings. This is crucial, since at present these pinched black holes
remain unknown. We shall develop this idea in the next subsection,
focussing first on the available information.

The behavior of non-uniform black
strings on $\CM^{n+2} \times S^1$ has been computed
numerically in \cite{Gubser:2001ac,Wiseman:2002zc,Sorkin:2004qq,Kudoh:2004hs,
Kleihaus:2006ee,Sorkin:2006wp}, and from the results in appendix \ref{app:torusphases}
it follows how to translate this into results for black membranes
with non-uniformity in one direction.
In particular, close to the GL point the non-uniform black membrane (nubm) has
\begin{equation}
\label{ahnum}
a_H^{\rm nubm} (\ell) = a_H^{\rm ubm}  (\ell) \left[ 1 - \frac{n^2 (n+1)}{2 (n-1)^2}
\frac{ \gamma_{n+2} }{\ell_{\rm GL}^{n+4}} ( \ell - \ell_{\rm GL})^2
+ \CO ( (\ell - \ell_{\rm GL})^3 ) \right]\,,
\end{equation}
where $ a_H^{\rm ubm}  (\ell) $ is the area function \eqref{aHmem} of the uniform
black membrane. Here, $\ell_{\rm GL} = (\mu_{{\rm GL},n+2})^{-\frac{1}{n+1}}$
is the critical GL wavelength in terms of the dimensionless GL mass
$\mu_{{\rm GL},d}$,
and $\gamma_{d}$ another numerically determined constant
(see tables 2 and 3 of \cite{Harmark:2007md} respectively
for the values of these for $ 4 \leq d \leq 14$). Also, when $\gamma  >0$
the non-uniform phase extends in the direction of $ \ell < \ell_{\rm
GL}$, while for $\gamma < 0$ it goes in the opposite direction,  $ \ell
> \ell_{\rm GL}$.

We can extract two important observations from this. First,
the result \eqref{ahnum} shows that the curve of the non-uniform phase
is tangent to the curve of the uniform phase at the GL point, since their entropies
differ only at second order away from this point. In fact, this can be
derived as a consequence of the Smarr relation and the first law \cite{Harmark:2007md}.
Second, the coefficient $\gamma_{d}$ is known to change \cite{Sorkin:2004qq} from positive to
negative for $d > 12$. Thus for $ 2 \leq n \leq 10$ the non-uniform phase extends to
$\ell - \ell_{\rm GL} < 0 $ and has less entropy than the uniform phase,
while for $n \geq 11$ the non-uniform phase extends to $\ell - \ell_{\rm GL} > 0 $
and has higher entropy than the uniform phase.

Another relevant aspect of phases that have $SO(n+1)$ symmetry and vary
along the circle direction, like the non-uniform and localized solutions, is
that one can easily
generate `copied' phases with multiple non-uniformity or
multiple localized black objects \cite{Horowitz:2002dc,Harmark:2003eg} (see also
\cite{Harmark:2003yz,Dias:2007hg}). Given the exact periodicity along
the circle, this is done by simply copying the
solution $k$ times on the circle. Clearly, this applies also to
the corresponding solutions on $\CM^{n+2} \times \T^2$ that we are
interested in. In the latter case,
for any solution with given $(\ell,a_H$) we get a new one with
\begin{equation}
\label{copy}
\tilde \ell = k^{\frac{n-1}{n+1}} \ell \spa
\tilde a_H = k^{- \frac{2}{n+1}} a_H\,.
\end{equation}
This applies in particular to the localized black string and non-uniform
black membrane phases, and, obviously,
leaves invariant the curve \eqref{aHmem} for the uniform black membrane.

\begin{figure}[ht!]
\begin{picture}(0,0)(0,0)
\put(300,35){\Large $\ell$} \put(45,270){\Large $a_H$}
\end{picture}
\centerline{\includegraphics[width=13cm]{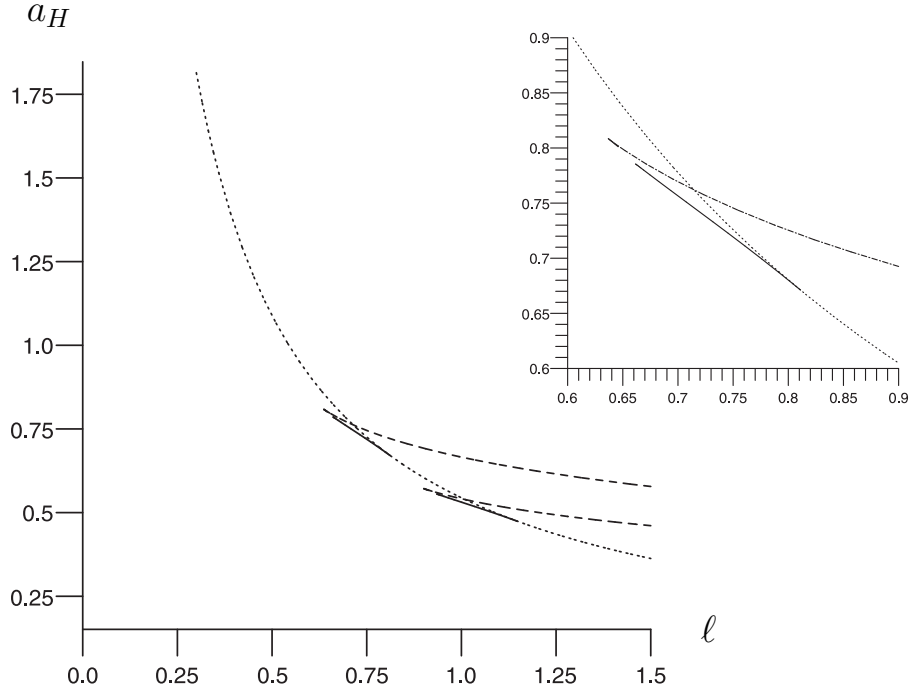}} \caption{
\small $a_H (\ell)$ phase diagram in seven dimensions ($\CM^5 \times
\T^2$) for Kaluza-Klein black hole phases with one uniform
direction. Shown are the uniform black membrane phase (dotted), the
non-uniform black membrane phase (solid) and the localized black
string phase (dashed). For the latter two phases, we have also shown
their $k=2$ copy. The non-uniform black membrane phase emanates from
the uniform black membrane phase at the GL point $\ell_{\rm GL} =
0.811 $, while the $k=2$ copy starts at the 2-copied GL point
$\ell_{\rm GL}^{(2)} = \sqrt{2} \ell_{\rm GL} =1.15 $. 
The connection between the curves is shown in greater detail.
The gap between the solid and dashed curves reflects the 
difficulty in getting numerical data close to the merger.
This figure is representative for
the phase diagram of phases on $\CM^{D-2} \times \T^2$ for all $ 6
\leq D \leq 14$. \label{fig:KKphases7}}
\end{figure}

Currently, the best available data for KK phases correspond to
six-dimensional KK black holes on $\CM^5 \times S^1$ (see \eg\
Ref.~\cite{Harmark:2007md} and references therein). We can use the
dictionary in appendix \ref{app:torusphases} to map the known data to
find the curves $a_H (\ell)$ for the corresponding phases on $\CM^5
\times \T^2$. The resulting phase diagram (based on the numerical
results of Ref.~\cite{Wiseman:2002zc,Kudoh:2004hs}) is depicted in
Fig.~\ref{fig:KKphases7}. It includes the uniform black membrane, the
black membrane with one uniform and one non-uniform direction, and the
black string localized in one of the circles of $\T^2$. We have also
included the $k=2$ copies obtained from these data and the map
\eqref{copy}. Both the uniform black membrane phase and
the localized black string phase extend to $\ell \rightarrow \infty$
where they obey the behavior \eqref{aHmem} and \eqref{aHstr}
respectively with $n=3$.

The seven-dimensional phase diagram displayed in
Fig.~\ref{fig:KKphases7} is believed to be representative for the black
membrane/localized black string phases on $\CM^{D-2} \times \T^2$ for
all $ 6 \leq D \leq 14$. Here the lower bound is obvious since the phase
diagram of KK black holes on $\CM^{D-2} \times S^1$ that we are
considering starts at $D=6$. On the other hand, the upper bound  follows
from the fact that, as explained above, there is a critical dimension
$D=14$ above which the behavior of the non-uniform black string phase is
qualitatively different \cite{Sorkin:2004qq}.

\begin{figure}[ht!]
\begin{picture}(0,0)(0,0)
\put(340,-5){$\ell$}
\put(95,238){$a_H$}
\end{picture}
\centerline{\includegraphics[width=8cm]{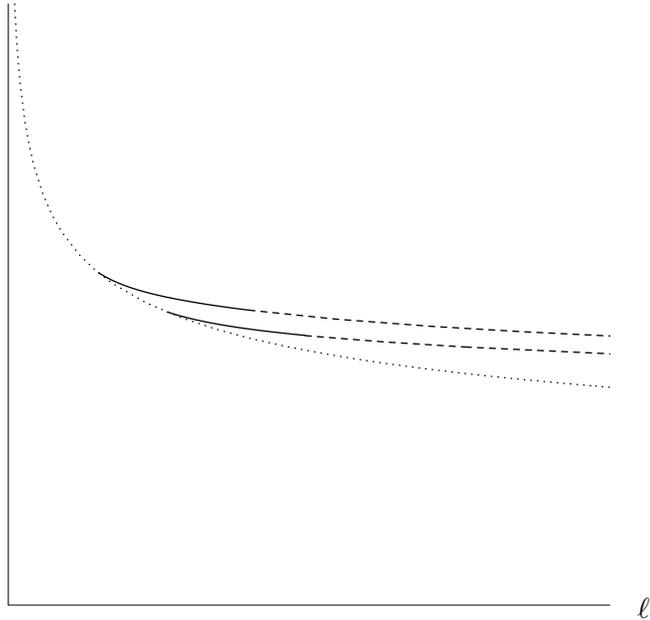}}
\caption{\small Expected $a_H (\ell)$ phase diagram for KK black hole
phases with one uniform direction on $\CM^{D-2}
\times \T^2$, when $D> 14$. We also show the $k=2$ copy of non-uniform
phases, obtained from the main sequence using \reef{copy}. The uniform
black membrane curve (dotted) and the asymptotic form of the localized
black string curves (dashed) are known exactly. Of the solid lines for
non-uniform black membranes, we only know the position of the GL points
where they begin, and the fact that they must be tangent to the uniform
membrane curve at this point. The points of merger to the black string
curves are unknown.
\label{fig:abovecrit}}
\end{figure}

The phase diagram for $D> 14$ is much poorly known in comparison,
since there are no data like fig.~\ref{fig:KKphases7} available for the
localized and non-uniform phases, only the asymptotic behaviors. We do
know, however, that the curve $a_H(\ell)$ for non-uniform membranes must
again be tangent to the curve of uniform membranes, but now it must lie
above the latter, \ie non-uniform branes have higher entropy than
uniform ones. A plausible form of the phase diagram, compatible with the
information that we have discussed, is presented in
fig.~\ref{fig:abovecrit}. Notice that as $n$ grows larger the asymptotic curves
for uniform black membranes and localized black strings, \reef{ajring} and
\reef{mpaj}, become flatter and closer. In
fig.~\ref{fig:abovecrit} these curves correspond to $D=14$ ($n=10$).

The phases discussed above are the presently known solutions on $\CM^{D-2}
\times \T^2$ that i)~have $SO(D-3)$ symmetry; ii)~have one uniform
direction; and iii)~are in thermal equilibrium. If one drops this last
condition, there are also multi-localized black string
solutions, arising from the multi-black hole configurations on
$\CM^{D-2} \times S^1$ obtained and studied in Ref.~\cite{Dias:2007hg}.
These do also have a counterpart for rotating black holes on $\CM^{D}$.

\subsection{Phase diagram of neutral rotating black holes on $\CM^D$
}
\label{subsec:conject}

The first observation based on the membrane analogy is that the
phase diagram of rotating black holes should also exhibit an
infinite sequence of lumpy (pinched) black holes emerging from the
curve of MP black holes at increasing values of $j$. This point was
made in \cite{Emparan:2003sy} but let us revisit and elaborate on it
a little more here.

It is easy to see, \eg\ by computing the ratio between the horizon curvatures
along the rotation plane and transverse to it, that at large $j$, \ie\
small $r_0/a$, the black membrane is a
good approximation to the MP horizon from the rotation axis
$\theta=0$ down to values of the polar angle of order
\beq\label{maxth}
\frac{\pi}{2}-\theta \sim \frac{r_0}{a}\,.
\eeq
Only very close to the equatorial edge is there a significant deviation
from the membrane geometry. For instance, the region $0\leq \theta
\lsim \frac{\pi}{2}- \frac{r_0}{a}$ covers almost all the
horizon area, and its length along $\theta$, up to terms of
order $r_0^2/a^2$.

Let us now import the GL zero-modes of the black membrane, including
the $k$-copies that appear at increasing $\ell$ according to
\reef{copy}, into the rotating black hole horizon.
We must choose axially-symmetric combinations of the zero modes, so we
change basis from plane waves $\exp(i k_{\rm GL} z)$ to Bessel
functions. Axially symmetric modes have a profile $J_0(k_{\rm
GL}a\sin\theta)$ \cite{Emparan:2003sy}. The basic and simple point here
is that the wavenumber, or wavelength $k_{\rm GL}^{-1}$, of the GL
zero-mode remains the same in the two analogue systems, to first
approximation, even if the profiles are not the same.

At large $j$ the wavelength of GL zero modes, $k_{\rm GL}^{-1}\sim r_0$,
is much smaller than the extent $\sim a$ along which the horizon looks
membrane-like, so we can fit many zeroes of $J_0(k_{\rm GL}a\sin\theta)$
in the horizon---this is the analogue of having a high-$k$-copy mode.
Over a length $\sim r_0$ on the horizon, the relative corrections to the
membrane metric due to finite rotation effects are of order
\cite{Emparan:2003sy}\footnote{This is the size of the corrections in
the off-diagonal terms. The corrections to diagonal terms are suppressed
by one more power of $r_0/a$.}
\beq\label{finitea}
\delta g_{\mu\nu} \sim \frac{r_0}{a}\;g_{\mu\nu}\,.
\eeq
So the profile of the zero mode will be approximately given by the Bessel
function down to polar angles \reef{maxth}, and only very near the
equator will there appear significant changes. This modifies the
boundary conditions along the horizon, relative to those for the black
membrane, but by locality, the behavior of modes with much smaller
wavelength than the horizon extent will not be significantly modified
by edge effects.

As a consequence we can predict the existence of an infinite sequence of
pinched black hole phases emanating from the MP curve at increasing
values $j_{\rm GL}^{(k)}$. The argument is clearly less strong for the
first few GL zero-modes, say, $k=1,2$, where $r_0$ and $a$ are
comparable so edge effects may become important. The case for the
existence of these pinched phases is more strongly made from the need to
complete the black ring and black Saturn curves as we shall see below.

We can make one more prediction for the further evolution of the new
branches away from the MP curve: like in the membrane case, the Smarr
relation and the first law can again be used to prove that two branches
of solutions coming out of the same point must be tangent in the
$(j,a_H)$ diagram. However, it is much more difficult to determine which
of the branches will have larger area. In particular, the corrections
computed in \reef{ahnum} enter at order $(r_0/L)^{n+4}$. But in the
rotating black hole, where we identify $L\sim a$, these are overwhelmed
by the finite-rotation corrections \reef{finitea}. So \reef{ahnum} can
not reliably be imported, and in particular the analogy does not allow
us to infer the existence of a critical dimension.

\subsubsection*{Main sequence}

\begin{figure}[th!]
\begin{picture}(0,0)(0,0)
\put(417,16){$j$}
\put(38,252){$a_H$}
\put(65,3){$j_{\rm mem}$}
\put(115,3){$j_{\rm GL}$}
\end{picture}
\centerline{\includegraphics[width=14cm]{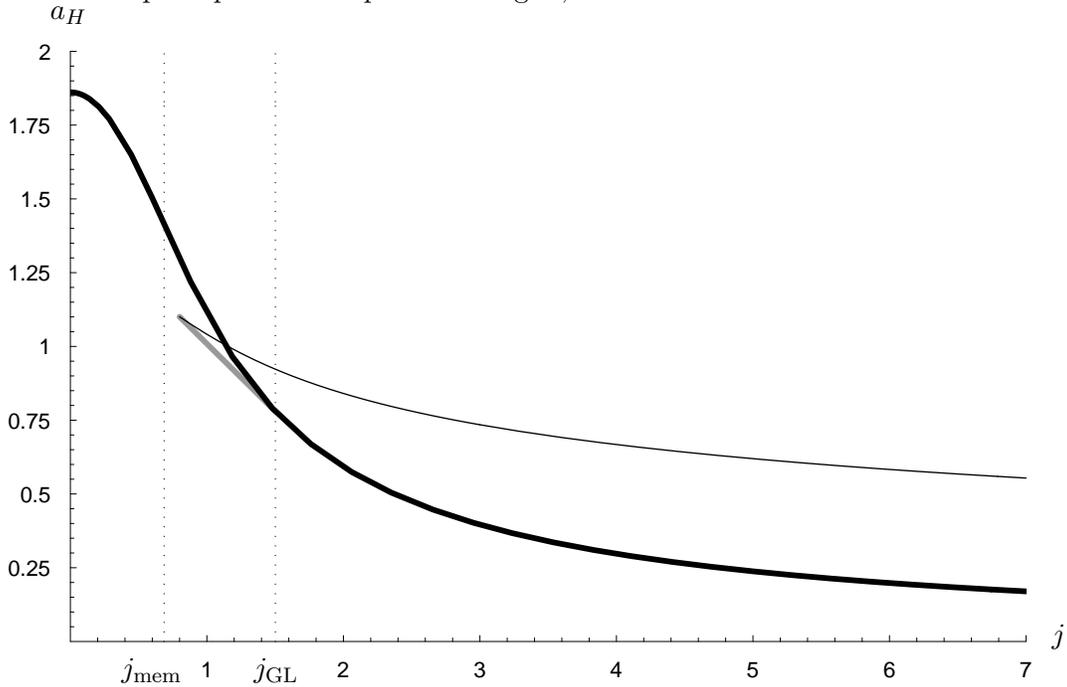}}
\caption{\small Qualitative completion of fig.~\ref{fig:MPBR} using
fig.~\ref{fig:KKphases7}. The gray line corresponds to the conjectured
phase of pinched black holes, which branch off tangentially from the MP
curve (thick) at a value $j_{\rm GL}> j_{\rm mem}$, and merge with the
black ring curve (thin). At any given dimension, the phases may not
display the swallowtail in fig.~\ref{fig:KKphases7}, depicted here, but
may instead be smoother like fig.~\ref{fig:abovecrit}. Even if there is
a cusp, the merger need not happen at it.}
\label{fig:sevendphases}
\end{figure}

The analogy developed above suggests that the phase diagram of
rotating black holes in the range $j>j_{\rm mem}$ where MP black holes
behave like black membranes, is qualitatively the same as
fig.~\ref{fig:KKphases7}, which describes actual membranes and other
phases in $\CM^{D-2} \times \T^2$. Thus we are naturally led to propose
that fig.~\ref{fig:MPBR} is completed to the phase diagram in
Fig.~\ref{fig:sevendphases}. At this moment we are only including the
main sequence of non-uniform phases, and not the `copied' phases also
present in fig.~\ref{fig:KKphases7}, which will be discussed later.

Several comments about this diagram are in order. The fact that
$a_H(j=0)$ is finite for MP black holes while $a_H(\ell=0)\to\infty$
for uniform black membranes, is inconsequential since these regions
lie at $j<j_{\rm mem}$ where the analogy breaks down. The figure
shows the pinched (lumpy) rotating black hole phase as a gray line
emerging tangentially from the MP black hole curve at a critical
value $j_{\rm GL}$ that is currently unknown. Arguments were given
in \cite{Emparan:2003sy} to the effect that $j_{\rm GL}\gsim j_{\rm
mem}$, consistently with the analogy. We may have the `swallowtail'
structure of first-order phase transitions (as depicted), or instead
that of second-order phase transitions, fig.~\ref{fig:abovecrit}.
But it may not be unreasonable to expect that a swallowtail appears
at least for the lowest dimensions $D=6,7,\dots$, since this is in
fact the same type of phase structure that appears for $D=5$.

As we move along the gray line in fig.~\ref{fig:sevendphases} in the
direction away from the MP curve, the pinch at the rotation axis of
these black holes grows deeper. Eventually, as depicted in
fig.~\ref{fig:membranes}, the horizon pinches down to zero thickness at
the axis and then the solutions connect to the black ring phase. For all
we know, if there is a cusp, the merger need not happen at it.

It would be interesting to study whether the merger happens through a
conical geometry, like
\cite{Kol:2002xz,Wiseman:2002ti,Kol:2003ja,Sorkin:2006wp} have argued
for non-uniform black string/membrane phases.

\subsubsection*{Black Saturns and multi-pinch sequences}

For KK phases in $\CM^{D-2} \times \T^2$ there are copies with multiple
non-uniformity. However, the implications of these configurations for
the phase diagram of rotating black holes on $\CM^{D}$ is not
straightforward since the analogy becomes less precise for them. The
reason is the following. The pinches on a non-uniform membrane are
related by a periodic translation symmetry, and so are all exactly
equivalent. However, there is no approximate translation symmetry for
the pinches on the rotating black hole horizon, even as the
ultra-spinning limit is approached. As we have seen, the profiles of the
zero-modes, \ie the pinches at linearized order, are approximately
Bessel functions $J_0(k_{GL} a\sin\theta)$ away from the equator. Since
these functions do not have any discrete translation symmetry along
$\theta$, there is actually no limit in which we recover exactly the
non-uniform copies for the system on $\T^2$. So even if the
multiply-pinched phases emerging from the MP curve are, by our arguments
above, a natural consequence of the analogy, their development further
away from the branch-off point cannot be deduced at all from the analogy.

Nevertheless, even if the correspondence breaks down, we can still
argue to infer some qualitative features. Let us proceed by
increasing the number of `copies'. As we have seen, the most precise
application of the analogy to non-uniform phases is to the main
sequence (no-copy) phases. For the system on $\T^2$ these start from
a GL perturbation of the membrane with a single minimum, and then
evolve as in fig.~\ref{fig:membranes}. Even if the analogue phases
(iv) are on one side straight and static, and on the other circular
and rotating, their curves \reef{aHrh} and \reef{aHstr} match up to
numerical factors.

Next, the first copy for the non-uniform membrane on $\T^2$ begins as a
GL zero-mode perturbation of the membrane with two minima, which grows
to merge with a configuration of two {\it identical} black strings
localized on the torus. For the MP black hole, the analogue is the
development of a circular pinch, which then grows deeper until the
merger with a black Saturn configuration in thermal equilibrium. Thermal
equilibrium, \ie\ equal temperature and angular velocity on all
disconnected components of the event horizon, is in fact naturally
expected for solutions that merge with pinched black holes, since the
temperature and angular velocity of the latter should be uniform on the
horizon all the way down to the merger, and we do not expect them to
jump discontinuously there. There is no reason to doubt the existence of
these black Saturns, but in contrast to the two strings on $\T^2$, the
two black objects in them are clearly not identical. Still, it is
possible to obtain a good approximation for black Saturns in the case
when the size of the central black hole is small compared to the radius
of the black ring, since then the interaction between the two objects is
small and, to a first approximation, one can simply combine them
linearly. In five dimensions, where a comparison with exact black Saturn
solutions is possible \cite{Elvang:2007rd}, this approximation has been
shown to be reliable \cite{Elvang:2007hg}. It can be readily extended to
any $D\geq 5$. One then easily sees that, under the assumption of equal
temperatures and angular velocities for the two black objects in the
black Saturn, as $j$ is increased a larger fraction of the total mass
and the total angular momentum is carried by the black ring, and less by
the central black hole. Then, this black Saturn curve must asymptote to
the curve of a single black ring.

Strikingly, there is a definite possibility that a second kind of black
Saturn, also in thermal equilibrium, exists in $D\geq 6$, which would
not have a counterpart in five dimensions. The reason is the following.
If we consider MP black holes, and instead of $M$ and $J$ we fix the
horizon temperature and angular velocity, then there exist two MP black
holes with those values of $T_H$ and $\Omega_H$\footnote{A maximum value
for $\Omega_H/T_H$ is attained at \reef{onset}.}: a black hole with a
rounded shape, and smaller values of both $M$ and $J$, and another one
with larger $M$ and $J$ and a pancaked shape. Now, in our phase
diagrams, we fix the total $M$ and $J$ of the black Saturn, but the mass
and spin of each of its two constituents are not fixed separately.
Instead, under thermal equilibrium we demand that the temperature and
angular velocity of the black ring and the central black hole are equal.
So besides the black Saturn with a small, round central black hole, that
we have discussed above, it may be possible to have another one with a
large, pancaked central black hole. These {\em pancaked black Saturns} do
not have a counterpart in five dimensions, since pancaked black holes
with large spin exist only in $D\geq 6$. To get a better idea of the
properties of these configurations, we may try to regard them as
made of a black ring and a MP black hole that satisfy
eqs.~\reef{tbrthermo} and \reef{TMP2}, \reef{JOm2}, respectively. Then,
if the temperatures and angular velocities of the two objects are equal,
they must have similar thickness $r_0$ and also similar extent along the
plane of rotation, $a\sim R$. The latter means that we cannot assume
that the two objects are far from each other and interact weakly, and so
a simple superposition never becomes really accurate. It is thus
conceivable that the black ring can not support itself under the
gravitational attraction of the central black hole and so these black
Saturns might not actually exist.

Nevertheless, let us assume that they do exist. In this case, even if
the approximation based on eqs.~\reef{tbrthermo} and \reef{TMP2},
\reef{JOm2} may not be too accurate, it still tells us that most of the
mass, angular momentum and area of the black Saturn are concentrated in
the central pancaked black hole, while much less of them (by a factor
$\sim r_0/a$) is in the very thin black ring around it. So the area
curve will asymptote (from below) to the curve of a single MP black
hole. The two black Saturn curves presumably bifurcate at a value $j\sim
O(1)$ near the point of merger with the pinched black hole, but this
could happen before, after, or at the merger point. It seems unlikely
that the pancaked black Saturn phase appears there and then joins back
at larger $j$ to the main-sequence MP curve, since this would appear to
require a second merger to a circularly-pinched black hole. Instead, it
is more plausible that the pancaked black Saturn continues to exist at
larger $j$. In this case, the pancaked central black hole will
presumably encounter a GL-like zero-mode from which a new branch of
solutions emerges, where the central black hole develops a pinch on the
rotation axis---so we will find a {\em pinched black
Saturn}\footnote{If, as we assume, the $S^{n+1}$ does not spin, a thin
black ring will not develop pinches. Whether this may happen at $j\sim
O(1)$ is not known.\label{fn:pinchring}}. We will discuss the possible
mergers of this phase after we encounter it again in our subsequent
discussion.

The analogy to non-uniform KK phases on $\T^2$ becomes even more
inadequate when we consider the next `copy', \ie one more pinch on the
rotating black hole horizon. It is certainly expected that at
sufficiently large $j$ the rotating MP black hole can fit a zero-mode GL
perturbation with both a central pinch and a circular pinch---so a
doubly-pinched phase emanates from the curve of smooth MP black holes.
However, whereas the pinches on a non-uniform membrane on $\T^2$ all
evolve, by symmetry, at exactly the same rate, this will not be the case
for the corresponding doubly-pinched black hole, where there is no
translation symmetry along the polar direction on the horizon. It is
therefore unlikely that both pinch down to zero simultaneously. We may
expect that the outer, circular pinch, grows deeper faster (in phase
space) than the central pinch, since the horizon is thinner away from
the axis. In this case, if pinch-down to zero occurs, the pinched black
hole will connect not to a black di-ring, but to a black Saturn! This
provides, then, a natural way to connect to the pinched black Saturns
discussed above\footnote{If instead the central pinch shrank faster than
the circular one, we would seem to connect to a pinched black ring. Again, this
is conceivable at $j\sim O(1)$, but it appears to require more complicated connections
to complete the phase diagram.}. If so, this would prevent the
pinch-down to zero of the second, central pinch, and so there would not
be any merger to a di-ring. Under these assumptions, there does not
appear to be any need, nor in fact much room in the phase diagram, for a
black di-ring phase (the `analogues' of the second-copy of localized
phases on $\T^2$) that, coming from the merger to a single-horizon
phase, would be in thermal equilibrium.

In fact the existence, as asymptotic curves at large $j$, of black
di-rings, and in general of multi-black rings in thermal equilibrium,
already seems difficult for the same reason as in five-dimensions
\cite{Elvang:2007hg}: fixing both the angular velocity and the
temperature uniquely determine a black ring, so two black rings in
thermal equilibrium with each other have the same radius and would
actually be on top of each other\footnote{Two thin black rings very
close to each other and in thermal equilibrium would appear to require a
merger to a thin pinched black ring, which (see footnote
\ref{fn:pinchring}) is unlikely.}. According to this, at least at large
values of $j$ multi-rings are unlikely (at smaller $j$ strong
gravitational effects might introduce changes). This is nicely
consistent with our conclusion above that the `second-copy' pinched
black hole does not merge with a di-ring phase but with another phase,
presumably a pinched black Saturn. The story repeats, with multiplied
complication, for the subsequent multiply-pinched phases.

\subsubsection*{Summary}

\begin{figure}[th!]
\centerline{\includegraphics[width=12cm]{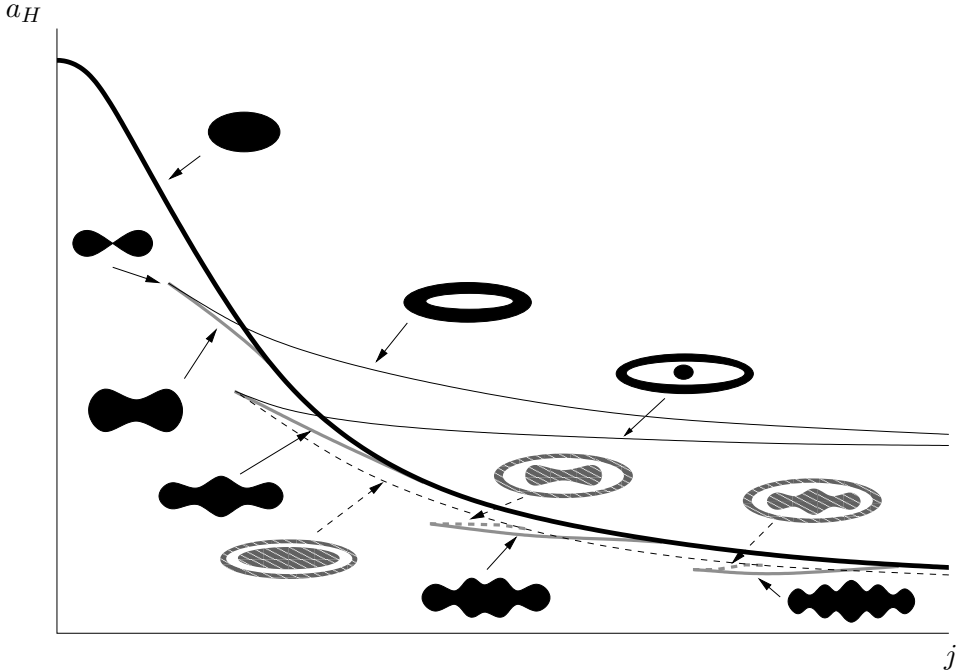}}
\begin{picture}(0,0)(0,0)
\put(25,250){$a_H$}
\put(380,5){$j$}
\end{picture}
\caption{\small Proposal for the phase diagram of thermal equilibrium phases
in $D\geq 6$. The solid lines and figures have significant arguments in
their favor, while the dashed lines and figures might not exist and
admit conceivable, but more complicated, alternatives. Some
features have been drawn arbitrarily:
at any given bifurcation and in any dimension, smooth connections like
fig.~\ref{fig:abovecrit} are possible instead of swallowtails with cusps; also, the
bifurcation into two black Saturn phases may happen before, after, or
right at the merger with the pinched black hole. Mergers to di-rings or
multi-ring configurations that extend to asymptotically large $j$ seem
unlikely. If thermal equilibrium is not imposed, the whole semi-infinite
strip $0<a_H <a_H(j=0)$, $0\leq j<\infty$ is covered, and multi-rings
are possible.}
\label{fig:hidphases}
\end{figure}

Fig.~\ref{fig:hidphases} gives a pictorial description of
the phase diagram that we propose. The features that we
believe can be argued with confidence (represented with solid lines and
figures) are:
\begin{enumerate}

\item The black ring phase and its merger to a phase of black holes with a
pinch at the rotation axis.

\item The upper black Saturn curve and its merger to a phase of black holes
with a circular pinch.

\item An infinite sequence of pinched black hole phases emanating from the MP
curve.

\item Any two phases that come out of the same point in the diagram must
be tangent at that point. However, for any bifurcation and in any given
dimension we cannot predict which of the two phases has higher or lower
area near the bifurcation.

\end{enumerate}
The arguments for the second point are as sound as those leading to the
first one: the existence of black Saturns can be easily argued when the
central black hole is small. The asymptotic value of their area at large
$j$ can then be determined and proven to asymptote from below to the
black ring value. The merger of these black Saturns to a pinched black
hole is then natural.

By constrast, the arguments for pancaked and pinched black Saturns, and
the merger of the latter to the pinched black holes that branch off the
MP curve, are comparatively less compelling and presumably admit
alternatives. Nevertheless, these conjectural phases provide a simple
and natural way of completing the curves in the phase diagram that is
consistent with all the information available.

We do not expect, at least at asymptotically large $j$, the existence of
multi-black rings, with or without a central black hole, in thermal
equilibrium. Then, at asymptotically infinite $j$ the only possible
phases in thermal equilibrium seem to be MP black holes, black rings,
and black Saturns (possibly in two varieties) with a single ring.

We stress that in the diagram of fig.~\ref{fig:hidphases} we have
represented, of all possible multi-black hole phases, only those that
can be in thermal equilibrium, \ie where the disconnected components of
the horizon all have the same temperature and angular velocity. This is
natural to impose for phases where disconnected components of the
horizon merge to a connected horizon.

In general, however, we expect the
existence of multi-black ring configurations, possibly with a central
black hole, in which the different black objects have in general
different surface gravities and different angular velocities%
\footnote{These configurations are the analogue of the multi-localized
string configurations on the torus that can be obtained from
multi-black hole configurations on the circle \cite{Dias:2007hg} by
adding a uniform direction.}%
. Although
they can not be in thermal equilibrium, they are perfectly valid as
stationary multi-black hole configurations of General Relativity---the
same remarks as in five dimensions \cite{Elvang:2007hg} apply to this
effect. Since they have additional parameters, they furnish continuous
families of solutions filling up a semi-infinite strip of the plane
$(j,a_H)$. At any non-zero $j$, the maximum area is achieved by black
Saturns with an almost static central black hole that carries most of the
mass, and a very thin and long black ring that carries most of the
angular momentum. Imposing thermal equilibrium eliminates the continuous
degeneracies and yields solutions characterized by curves $a_H(j)$ like
in fig.~\ref{fig:hidphases}.

\section{Discussion}
\label{sec:discuss}

The picture we have unveiled for phases of asymptotically flat
stationary black holes in higher dimensions is in stark contrast with
the four-dimensional phase diagram, which contains only the Kerr
solution. The five-dimensional phase diagram for thermal equilibrium
phases, fig.~2 in \cite{Elvang:2007hg}, exhibits some similarities to
the conjectured phase diagram for thermal equilibrium phases depicted in
fig.~\ref{fig:hidphases}, but there are several significant differences.
In five dimensions the merger between the MP and black ring curves
occurs at a naked singularity with zero-area, whereas in $D\geq 6$ the
pinched-down solution at the merger has finite area and the curvature
remains finite on the horizon away from the pinch. It is also amusing to
observe that the `fat black ring' branch that is present in $D=5$ is
replaced (at least to a large extent) by pinched black holes in $D\geq
6$. This is in accord with the fact that black membranes, uniform or
not, do not exist in $D=5$, and with the observation in
\cite{Elvang:2006dd} that fat black rings behave in many respects like
`drilled-through' black holes, rather than as circular black strings.
More strikingly, while in five dimensions there are only three kinds of
single-black hole phases---the MP black hole and the thin and fat black
rings---, in $D\geq 6$, following \cite{Emparan:2003sy}, we have been
led to include an infinite number of black holes with spherical topology
whose horizons have multiple concentric pinches. When they have more
than one pinch, their evolution in phase space and their mergers with
other thermal equilibrium phases become more speculative. We have also
pointed out the possibility of two kinds of thermal-equilibrium black
Saturns, only one of which has a five-dimensional counterpart.

So far we have not made almost any mention of the stability of the
new solutions, although much of the rich phase structure we have
found is actually due to the onset of horizon instabilities. Some
instabilities, however, are not expected to give rise to new phases.
For instance, black rings at large $j$ in any $D\geq 5$ are expected
to suffer from a GL instability that creates ripples along the $S^1$
and presumably fragments the black ring into black holes flying
apart \cite{Emparan:2001wn,Hovdebo:2006jy,Elvang:2006dd}. It is not
known whether this instability switches off or not at $j\sim O(1)$.
For large enough $j$, MP and pinched black holes in any dimension
might suffer from a similar instability creating ripples along the
rotation direction. This instability might radiate away the `excess'
angular momentum, or perhaps more likely, break the horizon apart
\cite{Emparan:2003sy}. Additionally, if turning points of $j$ appear
in the phase curve (such as the conjectural cusps in
fig.~\ref{fig:hidphases}) then, in analogy to the five-dimensional
case \cite{Arcioni:2004ww,Elvang:2006dd}, we may expect, for the two
branches of solutions meeting at the turning point, that the branch
with the lowest area will be unstable under radial perturbations to
collapse into an MP black hole. Typically, this would apply to
pinched black holes, although perhaps not to all of them. If such
turning-points for $j$ are absent, pinched black holes will
presumably be stable to radial perturbations.

There are several natural extensions of the present work. We discuss
some of them here.

A straightforward application of our methods is to study thin black
rings in external gravitational potentials. For instance, we may
study a thin black ring surrounding a central black hole---thus
yielding a black Saturn---, or a black ring in the presence of a
cosmological constant---\ie a black ring in AdS or dS spacetime.%
\footnote{In \cite{Kunduri:2006uh} the existence of supersymmetric
black rings in AdS is considered.} The latter may have interesting
implications in the AdS/CFT correspondence (see below for additional
comments on this). Black rings with charges
\cite{Elvang:2003yy,Elvang:2004rt} and with dipoles
\cite{Emparan:2004wy} also satisfy the zero-pressure condition
\reef{notzz} and can be analyzed with our methods.\footnote{The existence 
of small supersymmetric black rings in $D\geq 5$ is argued in
\cite{Dabholkar:2006za}.} 
These possibilities are currently under investigation.

The general ideas in sec.~\ref{sec:thincircle} as well as the specific
construction of the linearized solution in sec.~\ref{sec:equil} can in
principle also be extended to analyze black rings with horizon $S^1\times
S^{n+1}$ with rotation not only along $S^1$ but also in the $S^{n+1}$.
These involve more functions in the linearized solution, but no new
conceptual difficulty is envisaged. The main obstacle is that the
near-horizon perturbations of a black string with a rotating $S^{n+1}$
may be quite complicated and perhaps intractable analytically for $n>1$.
Nevertheless, we can anticipate that the rotation in the $S^{n+1}$ will
introduce particularly rich dynamics for $n\geq 3$, since it is then
possible to have ultraspinning regimes for this rotation too. In this case the
$S^{n+1}$ can also develop pinches and presumably connect to phases with
horizon $S^1\times S^1\times S^{n}$---and if $n\geq 4$ we can then
repeat the story for this last $S^n$. For six-dimensional black rings
$S^1\times S^{3}$, the $S^3$ can not pinch, but we may imagine
replacing it with a five-dimensional black ring and thus find a horizon $S^1\times
S^1\times S^{2}$.

Clearly, a daunting host of new possibilities opens up as we go
higher in $D$. It seems possible to extend the basic, simple core of
our ideas in sec.~\ref{sec:thincircle}, to obtain information about
the possible existence of more general {\it blackfolds}, obtained by
taking a black $p$-brane with horizon topology $\bbr{p} \times
S^{q}$ and bending $\bbr{p}$ to form some compact manifold. One must
then find out under which conditions a curved black $p$-brane can
satisfy the equilibrium equation \reef{KT}. Conventional approaches
based on topological considerations have only found very weak
restrictions in six or more dimensions
\cite{Helfgott:2005jn,Galloway:2005mf}, and, not being constructive,
provide scant information about the actual existence of horizons
with other topologies. On the other hand, it is possible to
construct time-symmetric initial data containing apparent horizons
with the geometry of products of spheres \cite{Schwartz:2007gj}, but
given the absence of rotation, dynamical evolution should presumably
drive these geometries to collapse into a single spherical black
hole. Our method, instead, is constructive and uses crucially
dynamical information to determine the possible horizon geometries.
Notice that several qualitatively new issues must be addressed for
$p>1$, since not only the topology, but also the embedding geometry
of the worldvolume of a thin black $p$-brane admits much richer
possibilities than in the case of a circular ring.

An alternative approach to the matched asymptotic expansion has been
developed in \cite{Chu:2006ce}. This is a systematic low-energy
(long-distance) effective expansion which gives results only in the
region away from the black hole and so it does never give directly the
corrections to near-horizon magnitudes, only to asymptotic magnitudes.
But one can again use Smarr relations and the first law, this time to
obtain the corrected area, temperature, and angular velocity. It would
be interesting to develop this method to describe extended brane-like
black holes such as black strings and possibly other blackfolds.

Another more indirect approach to higher-dimensional black rings in AdS
has been put forward recently by Lahiri and Minwalla
\cite{Lahiri:2007ae}. In a clever use of the AdS/CFT correspondence they
studied stationary, axially symmetric spinning configurations of plasma
in ${\cal N}=4$ SYM theory compactified to $d=3$ on a Scherk-Schwarz
circle. Such configurations can be studied with the use of the
relativistic Navier-Stokes equations, which describe the dynamics of the
dual of the horizon, while the radial coordinate away from the horizon
is holographically encoded. Thus, one works with equations that depend
on one less coordinate than in the gravitational dual, and this makes
them more easily tractable. The solutions to these equations correspond
to large rotating black holes and black rings in the dual Scherk-Schwarz
compactified AdS$_5$ space. Impressively, the phase diagram of these
rotating fluid configurations, even if dual to black holes larger than
the AdS radius, reproduces many of the qualitative features of the MP
black holes and black rings in five-dimensional flat spacetime.
Additional features like the existence of a maximum angular momentum for
large black rings presumably reflect the fact that AdS acts like a
gravitational well opposing the growth of black rings.
Higher-dimensional generalizations of this setup give predictions for
the phases of black holes in Scherk-Schwarz compactified AdS$_D$ with
$D>5$. Thus, ref.~\cite{Lahiri:2007ae} found stationary rotating
configurations of fluid that predict rotating black rings in AdS$_6$ and
`pinched' black holes that seem to support, in the AdS context, the
conjecture of `lumpy' black holes originally made in
\cite{Emparan:2003sy} and which we have found necessary for completing
the higher-dimensional phase diagram. AdS might limit the number of
possible pinches and so only a finite number of pinched plasma balls might
exist in this context.

The exploration of the possible blackfolds in terms of exact
analytic solutions, which has been very successful in four and five
dimensions, may turn out to be a hopeless task in higher-dimensional
spaces. We hope that the present paper helps to stimulate new
approaches to progress further into this fascinating subject.

\medskip
\section*{Acknowledgements}
\noindent This work was begun at the KITP, Santa Barbara, during the
program ``Scanning new horizons: GR beyond 4 dimensions", Jan-Mar
2006, and then continued at the workshop ``Einstein's Gravity in
Higher Dimensions", Jerusalem Feb.~18-22, 2007, the program ``String
and M theory approaches to Particle Physics and Cosmology", Galileo
Galilei Institute, Florence, Spring 2007, and the workshop
``Pre-Strings 2007", Granada, June 18-22, 2007. We are very grateful
to the organizers of these estimulating conferences, and to many of
the participants, in particular Oscar Dias, Henriette Elvang, Gary
Horowitz, Veronika Hubeny, Barak Kol, Rob Myers, Mukund Rangamani,
Amitabh Virmani, and Toby Wiseman for many useful discussions. RE is
also grateful to Filippo Brunelleschi from Florence for his
inspiration in solving the problem of mechanical equilibrium of
gravitating curved branes. All the authors acknowledge support by
the European Community FP6 program MRTN-CT-2004-005104. RE was
supported in part by DURSI 2005 SGR 00082, CICYT FPA
2004-04582-C02-02. TH would like to thank the Carlsberg Foundation
for support. VN acknowledges partial financial support by the INTAS
grant, 03-51-6346, CNRS PICS \#~2530, 3059 and 3747 and by the EU
under the contracts MEXT-CT-2003-509661, and MRTN-CT-2004-503369.
MJR was supported in part by an FI scholarship from Generalitat de
Catalunya.

\newpage

\section*{Appendices}

\begin{appendix}

\section{Ring-adapted coordinates for flat space}
\label{sec:flatcoords}

In this appendix we present a useful set of coordinate systems
in flat space suitably adapted to a circular ring.
We begin by writing the metric of Euclidean flat space $\bbe{n+3}$
as in \reef{bipolar}.
A hypothetical ring located at $r_1 = 0$ and $r_2 = R$ is an
$S^1$ curve embedded in $\bbe{n+3}$.

Let us consider now a point $\cal P$ located at radii $r_1$ and $r_2$,
angle $\psi=0$ in $S^1$ and arbitrary angular location in $S^{n}$. The distance
between $\cal P$ and the generic ring point at $r_1=0$, $r_2=R$ and
angle $\psi$ is
\begin{equation}
\label{appbb}
L(r_1,r_2,\psi) = \sqrt{r_1^2 + (R\cos \psi - r_2)^2 + R^2 \sin^2
\psi }\,.
\end{equation}
Then $\nabla^2 \left(L(r_1,r_2,\psi)^{-(n+1)}\right)=0$, with a delta-source at the ring
point.
Integrating over $\psi$ we obtain the total scalar potential at $\cal P$
\begin{equation}
\label{appbc}
\Sigma(r_1,r_2) = \frac{1}{2\pi}\int_{\psi=0}^{2\pi}
L(r_1,r_2,\psi)^{-(n+1)}\,
\end{equation}
such that $\nabla^2 \Sigma=0$ with distributional sources on the ring.
To obtain a coordinate system adapted on the equipotential surfaces of
$\Sigma$ we define the coordinates $\rho$ and $u$ by
\begin{equation}
\label{appbd}
\rho^{n+1} = \frac{1}{\Sigma(r_1,r_2)}
~,
\end{equation}
\begin{equation}
\label{appbe}
\partial_{r_1} u = - r_1^{n} r_2 \partial_{r_2} \Sigma
\spa \partial_{r_2} u = r_1^{n} r_2 \partial_{r_1} \Sigma
~.
\end{equation}
The definition of $u$ is such that it is orthogonal to $\rho$, \ie
$g_{\rho u}=0$.
In these coordinates the metric is
\begin{equation}
\label{appbf}
ds^2(\bbe{n+3}) = g_{\rho\rho} d\rho^2 + g_{uu} du^2 + r_1^2
d\Omega_{n}^2 + r_2^2 d\psi^2
\end{equation}
with
\begin{equation}
\label{appbg}
g_{\rho\rho} = \frac{(n+1)^2
\Sigma^{\frac{2(n+2)}{n+1}}}{(\partial_{r_1}
\Sigma)^2+(\partial_{r_2} \Sigma)^2} \spa g_{uu} =
\frac{1}{r_1^{2n} r_2^2 \left[(\partial_{r_1}
\Sigma)^2+(\partial_{r_2} \Sigma)^2 \right]}
~.
\end{equation}

For the considerations of the main text it is useful to distinguish
between the following regions of spacetime.

\subsubsection*{The far-zone: $r_1 , r_2 \gg R$}

The leading zeroth order behavior of the potential $\Sigma$
in the asymptotic region $r_1,r_2 \gg R$ is
\begin{equation}
\label{appbi}
\Sigma(r_1,r_2) = (r_1^2+r_2^2)^{-\frac{n+1}{2}}
~.
\end{equation}
This implies
\begin{equation}
\label{appbj}
\rho = \sqrt{r_1^2+r_2^2} \spa u =
\frac{r_1^{n+1}}{(r_1^2+r_2^2)^{\frac{n+1}{2}}}
~.
\end{equation}
Defining an angle $\Theta$ such that $u=(\cos \Theta)^{n+1}$ we get
\begin{equation}
\label{appbk}
r_1 = \rho \cos \Theta \spa r_2 = \rho \sin \Theta
\end{equation}
and
\begin{equation}
\label{appbl}
ds^2(\bbe{n+3}) = d\rho^2 + \rho^2 d\Theta^2 + \rho^2 \cos^2 \Theta d\Omega_{n}^2 +
\rho^2 \sin^2 \Theta d\psi^2
~.
\end{equation}

\subsubsection*{The near-ring-zone: $r_1 \ll R$ and $|r_2 - R| \ll R$}

In the near-ring region
it is more convenient to define a new set of
coordinates $r$ and $\theta$ as
\begin{equation}
\label{appca}
r(\rho)^{n} = \frac{k_n \rho^{n+1}}{R} \spa u(\theta) =
\mbox{const} + n k_n \int_{\theta'=0}^{\theta} d\theta'
(\sin \theta')^{n}
\end{equation}
with
\begin{equation}
\label{appcb}
k_n  \equiv \frac{\Omega_n}{2\Omega_{n-1}}=
\frac{\Gamma ( \frac{n}{2})}{2\sqrt{\pi} \Gamma (\frac{n+1}{2})}
~.
\end{equation}
To focus locally around the ring we take the limit
\begin{equation}
\label{appcc}
R \rightarrow \infty \spa r_1 \ \mbox{fixed} \spa {\tilde r_2} = r_2 - R
 \ \mbox{fixed} \spa z = R \psi  \ \mbox{fixed}
\end{equation}
and expand $\Sigma$ up to first order in $1/R$.
Since
\begin{equation}
\label{appcd}
L^{-(n+1)} = \frac{1}{(r_1^2+{\tilde r_2}^2+z^2)^{\frac{n+1}{2}}} - \frac{n+1}{2R}
\frac{{\tilde r_2} z^2}{(r_1^2+{\tilde r_2}^2+z^2)^{\frac{n+3}{2}}}
\end{equation}
we deduce that
\begin{equation}
\label{appce}
\Sigma = \frac{k_n}{R (r_1^2+{\tilde r_2}^2)^{\frac{n}{2}}}
\left(1-\frac{{\tilde r_2}}{2R}\right)
~.
\end{equation}
Then
\begin{equation}
\label{appcf}
r=\sqrt{r_1^2+{\tilde r_2}^2} \left( 1 + \frac{{\tilde r_2}}{2nR} \right)
~,
\end{equation}
\begin{equation}
\label{appcg}
u = k_{n+2} \left( \frac{r_1}{{\tilde r_2}} \right)^{n+1} ~
_2F_1 \left(\frac{n+1}{2},\frac{n+2}{2}; \frac{n+3}{2} ; -\frac{r_1^2}{{\tilde r_2}^2}
\right) +  \frac{k_n}{2 R} \frac{r_1^{n+1}}{(r_1^2 +{\tilde r_2}^2)^{\frac{n}{2}}}
~,
\end{equation}
\begin{equation}
\label{appci}
\cos \theta = \frac{{\tilde r_2}}{\sqrt{r_1^2+{\tilde r_2}^2}} \left( 1 -
\frac{1}{2n} \frac{r_1^2}{{\tilde r_2} R} \right)
~.
\end{equation}
The inverse coordinate transformations are
\begin{equation}
\label{coordtrans}
r_1 =r \sin \theta -
\frac{r^2}{2nR}\sin 2\theta\,,
\qquad
r_2={\tilde r_2}+R= R+r  \cos \theta -
\frac{r^2}{2nR}\cos 2\theta  ~.
\end{equation}
Plugging these expressions into the metric
we find the leading $1/R$ form
\begin{equation}
\label{appcj}
ds^2(\bbe{n+3}) = \left( 1 + \frac{2r\cos \theta}{R} \right) dz^2
+
\left( 1 - \frac{2}{n} \frac{r}{R} \cos \theta \right)
\left( dr^2 + r^2 d\theta^2 + r^2 \sin^2 \theta d\Omega_{n}^2
\right)
~.
\end{equation}
This metric coincides with
eq.\ \eqref{adapted}, where the adapted coordinates in the near-ring zone were
deduced with a more direct approach. The method used here allows us to
relate the near-ring coordinates $(r,\theta)$ to the coordinates $(r_1,r_2)$ valid
everywhere.

\section{Relations among the $f_i$'s}
\label{app:fis}
Given the three sources $T_{tt}$, $T_{t\psi}$, $T_{\psi\psi}$ and the
asymptotic boundary conditions, the solution to the linearized Einstein
equations is determined, up to gauge transformations, by three
independent functions. In \reef{hone} we introduced
six functions $f_i$. Here we explain how we can easily derive one of the
relations that exists among them.

To find this relation, begin by considering the problem in Cartesian coordinates in
the complete, asymptotically flat spacetime (\ie not just in the overlap
zone). We have
\begin{subequations}
\beqa
\nabla^2 \bar h_{tt}&=&-16\pi G T_{tt}\,,\\
\nabla^2 \bar h_{t i}&=&-16\pi G T_{t i}\,,\\
\nabla^2 \bar h_{ij}&=&-16\pi G T_{ij}\,,
\eeqa
\end{subequations}
where the $T_{ti}$ are obtained from $T_{t\psi}$ and the
$T_{ij}$ from $T_{\psi\psi}$ by simply changing coordinates.
Now pass from Cartesian to bi-polar coordinates \reef{bipolar}. Again we
will have several non-vanishing components $\bar h_{\mu\nu}$.
Notice, however, that
\beq\label{homom}
\bar h_{\Omega\Omega}=0\,
\eeq
since it is not sourced by any component of the stress tensor. If we now
effect the passage to adapted coordinates \reef{adapted}, the components
of the metric correction will get mixed in a rather complicated fashion,
but \reef{homom} will still hold.
Going over to
\beq
h_{\mu\nu}=\bar h_{\mu\nu}-\frac{1}{n+1}g_{\mu\nu}\bar h\,,
\eeq
all components $h_{\mu\nu}$, and hence all the $f_i$ in \reef{hone}, will be non-zero, but
\reef{homom} implies,
\beq
h_{\Omega\Omega}=\frac{1}{2}h g_{\Omega\Omega}\,.
\eeq
This immediately implies \reef{constr}.

\section{Regularity of the solutions}
\label{app:regular}

In this appendix we examine in detail the implications of regularity
for the functions $f_i^{(1)}$ that appear in the overlap-zone analysis
of sec.\ \ref{sec:solving}.

\subsection*{The equation for $f_1^{(1)}$ and $f_3^{(1)}$}

The $(tt)$ and $(zz)$ equations in sec.~\ref{sec:solving} take the
form (see \reef{rtt}), \beq f''+n\cot\theta\, f'-(n-1)f=0\,. \eeq
The change $y=(\sin\theta)^{\frac{n-1}{2}}f$ turns this equation
into an associated Legendre equation \beq y''+\cot\theta\, y'
+\left( \ell(\ell+1)-\frac{m^2}{\sin^2\theta}\right)y=0 \eeq (often
also written after setting $x=\cos\theta$) with indices
$\ell=\frac{n-3}{2},\; \frac{1-n}{2}$ and $m=\pm \frac{n-1}{2}$. The
two independent solutions can be taken to be the associated Legendre
functions of the first and second kind,
$P_{\frac{n-3}{2}}^{\frac{1-n}{2}}(\cos\theta)$ and
$Q_{\frac{n-3}{2}}^{\frac{n-1}{2}}(\cos\theta)$. So the general
solution of \reef{rtt} is \beq
f=(\sin\theta)^{\frac{1-n}{2}}\left(c_1
P_{\frac{n-3}{2}}^{\frac{1-n}{2}}(\cos\theta)+ c_2
Q_{\frac{n-3}{2}}^{\frac{n-1}{2}}(\cos\theta)\right)\,. \eeq The
prefactor $(\sin\theta)^{\frac{1-n}{2}}$ tends to introduce
singularities at $\theta=0,\pi$. The associated Legendre equation
has solutions that are non-singular on $\theta\in[0,\pi]$ iff
$\ell,m\in\Z$ with $0\leq m\leq \ell$ (or equivalent negative
values). So in general the solutions for $f$ are singular at both
$\theta=0,\pi$. Non-zero constants $c_1$ and $c_2$ can be chosen to
cancel the singularities at either $\theta=0$ or $\theta=\pi$, but
not at both. So regularity requires $c_1=c_2=0$, \textit{i.e.} \beq
f=0\,. \eeq

\subsection*{The equation for $f_6^{(1)}-f_5^{(1)}$}

Equation \reef{rth} is of the form \beq\label{rth2} f'+(n-1)\cot\theta
f-B \sin\theta=0 \eeq with $f=f_6^{(1)}-f_5^{(1)}$ and
$B=\frac{n+2}{n+1}\tau$. According to \reef{nocone} we must find
solutions such that
\beq\label{poles}
f(0)=f(\pi)=0\,.
\eeq

Defining $w=(\sin\theta)^{n-1}f$, eq.~\reef{rth2}
becomes
\beq
w'-B(\sin\theta)^{n}=0\,.
\eeq
The general solution of this equation can be written in terms of
the hypergeometric function $_2F_1(\alpha,\beta;\gamma;z)$,
\beq\label{uhyper}
w=k-B\cos\theta\; _2F_1 \left(\frac{1}{2},
\frac{1-n}{2};\frac{3}{2};\cos^2\theta\right)\,,
\eeq
with $k$ an integration constant.
Eq.~\reef{poles} requires that $w$ vanishes at $\theta=0$ and
$\theta=\pi$ faster than $\theta^{n-1}$ and $(\pi -\theta)^{n-1}$, resp.
Since $\gamma-\alpha-\beta=\frac{n+1}{2}>0$, the hypergeometric function is finite at
$z=\cos^2\theta=1$. Actually it takes the same finite value at $\theta=0$
and $\theta=\pi$, but then the factor $\cos\theta$ that
multiplies it in \reef{uhyper} makes it impossible to choose $k$ so
that $w$ vanishes both at $\theta=0$ and $\theta=\pi$.
So \reef{poles} can only be
achieved if $B=0$, \ie  if
\beq
\tau=0\,.
\eeq

\section{Solution for ${\sf F}$ and ${\sf G}'$}
\label{app:FandGp}

Given the solutions for ${\sf A}$ and ${\sf B}$, eqs.~\reef{usol} with
coefficients \reef{c3c4} and \reef{c1c2}, it is straightforward to plug
them into \reef{feq} and \reef{gpeq} to obtain ${\sf F}$ and ${\sf G}'$.
Computing the derivatives using
\beq
\frac{d}{d z}\; {}_2F_1(\alpha,\beta;\gamma;z)=
\frac{\alpha\beta}{\gamma}\;{}_2F_1(1+\alpha,1+\beta;1+\gamma;z)
\eeq
and working out the algebra we find
\beqa
\label{Fsol}
\frac{1}{A_1}{\sf F}&=&
-r\frac{  2 n^2
(n+2)-\left(3 n^3+4 n^2+n+4\right)
   \frac{r_0^{n}}{r^{n}}+2 (n+2) \frac{r_0^{2n}}{r^{2n}}}{n^2 (n+1)
\left(1-\frac{r_0^{n}}{r^{n}}\right)\frac{r_0^{n}}{r^{n}}}
\, _2F_1\left(-\frac{1}{n},-\frac{n+1}{n};1;1-
\frac{r_0^{n}}{r^{n}}\right)
\nn\\
&&- r\frac{4 (3 n^2 +6n+4)-\left(3 n (n+3)^2+20\right) \frac{r_0^{n}}{r^{n}}+4
   \frac{r_0^{2n}}{r^{2n}}}{n^2 (n+1)^2
   \left(1-\frac{r_0^{n}}{r^{n}}\right)}
\, _2F_1\left(-\frac{1}{n},\frac{n-1}{n};1;1-\frac{r_0^{n}}{r^{n}}\right)
\nn\\
&&- r\frac{2 \left(n^2-\frac{r_0^{n}}{r^{n}}\right)}{n^3}
\, _2F_1\left(-\frac{1}{n},\frac{n-1}{n};2;1-\frac{r_0^{n}}{r^{n}}\right)
\nn\\
&&+\frac{r_0^{n}}{r^{n-1}}\frac{2 (n-1) \left(3 n^2+6 n+4-\frac{r_0^{n}}{r^{n}}\right)}{n^3
(n+1)^2}
\, _2F_1\left(\frac{2n-1}{n},\frac{n-1}{n};2;1-\frac{r_0^{n}}{r^{n}}\right)
\,,
\eeqa
and
\beqa
\label{Gpsol}
\frac{1}{A_1}{\sf G}'&=&
\left(
\frac{(n+2) \left(1-2 n^2 \frac{r^{n}}{r_0^{n}}\right)}{n^2 (n+1)}-\frac{2 (n-1)}{n^2
   \left(1-\frac{r_0^{n}}{r^{n}}\right)}
\right)
\, _2F_1\left(-\frac{1}{n},-\frac{n+1}{n};1;1-\frac{r_0^{n}}{r^{n}}\right)
\nn\\
&&-\frac{4 \left(3 n^2+6 n+4\right)-\left(5 n^2+11 n+12\right)
   \frac{r_0^{n}}{r^{n}}+\left(2-n-n^2\right)
   \frac{r_0^{2n}}{r^{2n}}}{n^2 (n+1)^2
   \left(1-\frac{r_0^{n}}{r^{n}}\right)}
\, _2F_1\left(-\frac{1}{n},\frac{n-1}{n};1;1-\frac{r_0^{n}}{r^{n}}\right)
\nn\\
&&-\frac{2 n^2-(n+2) \frac{r_0^{n}}{r^{n}}}{n^3}
\, _2F_1\left(-\frac{1}{n},\frac{n-1}{n};2;1-\frac{r_0^{n}}{r^{n}}\right)
\nn\\
&&+\frac{r_0^{n}}{r^{n}}\frac{(n-1)  \left(6 n^2+12
   n+8-(n+2) \frac{r_0^{n}}{r^{n}}\right)}{n^3 (n+1)^2}
\, _2F_1\left(\frac{2n-1}{n},\frac{n-1}{n};2;1-\frac{r_0^{n}}{r^{n}}\right)
\,.
\eeqa
For clarity, we have factored out the overall coefficient $A_1$, whose value is given
in \reef{A1}.

\section{The five-dimensional black ring solution}
\label{app:5D}

Here we recover the known solution for a black ring in five
dimensions \cite{Emparan:2001wn,Emparan:2006mm},
in the limit of small $r_0/R$, using the
methods developed in this paper. This is of interest not only as a check
of our approach, but also because there is one significant difference
with respect to the higher-dimensional ($n>1$) cases that merits special
attention. It implies the existence of corrections to the physical
magnitudes $\mathcal{A}$, $M$, $J$, $\kappa$, $\Omega$ at order $1/R$.

\subsection{The linearized solution in `ring coordinates' $(x,y)$}

The five-dimensional black ring is most often written using a set of
adapted coordinates $(x,y)$, in which five-dimensional Minkowski space is
\beq\label{flatxy}
g^0_{\mu\nu}dx^\mu dx^\nu=-dt^2 +\frac{R^2}{(x-y)^2}\left[\frac{dy^2}{y^2-1}+
(y^2-1)d\psi^2+\frac{dx^2}{1-x^2}
+(1-x^2)d\phi^2\right]\,.
\eeq
See \cite{Emparan:2006mm} for an explanation of this coordinate system.
Here $-R/y$ plays the role of a radial coordinate and $\arccos (x)$ is a
polar angle, but constant $y$ and constant $x$ do not coincide with
constant $r$ and $\theta$ as we have defined them in this paper.

We solve the linear Einstein equations
with $T_{\mu\nu}$ given by
\beq\label{tmunu0}
T_{tt}=\frac{M}{2\pi R}\delta(-R/y)\,,\qquad
T_{t\psi}=\frac{J}{2\pi R}\delta(-R/y)\,.
\eeq
The ring lies at $y\to -\infty$.
The linearized Einstein equations in transverse gauge \reef{eins} are very easy to
solve for the sources \reef{tmunu0}. The $tt$
equation is the Laplace equation with sources
on a ring, whose solution is
\beq
\bar h_{tt}=\frac{2GM}{\pi R^2}(x-y)
\eeq
while the $t\psi$ equation is the same as we get for the $B_{t\psi}$
field of a ring of string \cite{Emparan:2006mm},
\beq
\bar h_{t\psi}=-\frac{2GJ}{\pi R^2}(1+y)\,.
\eeq
Then
$h_{\mu\nu}=\bar h_{\mu\nu}-\frac{1}{3}\bar h g_{\mu\nu}$ is
\beq
h_{tt}=\frac{4GM}{3\pi R^2}(x-y)\,,\quad h_{t\psi}=-\frac{2GJ}{\pi
R^2}(1+y)\,,\quad
h_{ij}=g_{ij}\frac{2GM}{3\pi R^2}(x-y)
\eeq
In principle, here $M$ and $J$ are independent quantities since their
respective sources enter separately the linearized equations, but we
know they must be related. We
introduce a dimensionless parameter
\beq
\nu=\frac{2GM}{3\pi R^2}=\frac{\sqrt{2}GJ}{\pi R^3}
\eeq
so
\beq
h_{tt}=2\nu(x-y)\,,\quad h_{t\psi}=-\sqrt{2}\nu R(1+y)\,,\quad
h_{ij}=\nu(x-y)g_{ij}^0
\eeq
This solution is not in the same gauge as in \cite{Emparan:2004wy,Emparan:2006mm}.
To go to the latter, change coordinates
\beq
x=\hat x-\frac{\nu}{2}(\hat x^2-1)\,,\qquad y=\hat y-\frac{\nu}{2}(\hat y^2-1)
\eeq
so that
\beq
g_{x x}d x^2=(1+\nu \hat y)\hat g_{\hat x\hat x}d\hat x^2   \,,\qquad
g_{\phi\phi}= (1+\nu \hat y)\hat g_{\phi\phi}
\eeq
and correspondingly for $x\leftrightarrow y$, $\phi\to \psi$. The
new solution, dropping hats, is
\begin{subequations}
\beq
h_{tt}=2\nu(x-y)\,,\quad h_{t\psi}=-\sqrt{2}\nu R(1+y)\,,
\eeq
\beq
h_{\psi\psi}=(\nu x+\nu(x-y))g^0_{\psi\psi}\,,\quad
h_{\phi\phi}=(\nu y+\nu(x-y))g^0_{\phi\phi}\,,
\eeq
\beq
h_{xx}=(\nu y+\nu(x-y))g^0_{xx}\,,\quad
h_{yy}=(\nu x+\nu(x-y))g^0_{yy}\,.
\eeq
\end{subequations}
This reproduces the solution in \cite{Emparan:2004wy,Emparan:2006mm}
to linear order in $\nu$.

\subsection{The solution in $(r_1,r_2)$ and $(r,\theta)$ coordinates:
the issue of $1/R$ corrections}

The previous calculation showed how the known solution is recovered to
linearized approximation around flat space. However, a puzzling fact
emerges when one notices that the expressions for $\mathcal{A}$, $M$,
and $J$ in the exact solution, when expanded in $\nu=r_0/R$, receive
corrections at linear order. This would seem to contradict our general
argument in section~\ref{sec:nearhor} that, since the perturbations to the
boosted black string are of dipole type, there should not be any
corrections. The resolution is instructive.

The integrals \reef{Phisol}, \reef{Asol} for the linearized
potentials can be
calculated explicitly (see \cite{Emparan:2001ux}),
\begin{subequations}\label{PhiA}
\beqa
\Phi&=&\frac{4 GM}{3\pi}\frac{1}{\sqrt{\left(r_1^2+r_2^2+R^2\right)^2-4r_2^2 R^2}}\,,\\
A&=&\frac{2GJ}{\pi R^2}
\left(\frac{r_1^2+r_2^2+R^2}{\sqrt{\left(r_1^2+r_2^2+R^2\right)^2-4r_2^2 R^2}}-1\right)\,.
\eeqa
\end{subequations}
The
relation between the coordinates $(r_1,r_2)$ and $(x,y)$ used in the
previous section, can be found in \cite{Emparan:2006mm}. It is then
straightforward to check that \reef{PhiA} yield the same
solution (in transverse gauge) as the linearized solution of the
previous section. This solution can be expanded now in the overlap zone
$r_0\ll r\ll R$, using the change of coordinates
\reef{coordtrans} and the relation \reef{MJA} between parameters. We find
\beqa\label{PhiAover}
\Phi&=&\frac{r_0}{r}\left(1+O\left(\frac{r_0}{r}\right)+O\left(\frac{r^2}{R^2}\right)\right)\,,\\
A&=&R\frac{\sqrt{2}r_0}{r}\left(1+\frac{r(\cos\theta-
1)}{R}+O\left(\frac{r_0}{r}\right)+O\left(\frac{r^2}{R^2}\right)\right)\,.
\eeqa
Then, to the appropriate order in the overlap zone,
\begin{subequations}\label{gmunu2}
\beqa\label{gtt}
g_{tt} &=& - 1 + \frac{2r_0}{r} \,,\\
\label{gtz}
g_{tz} &=& -\frac{\sqrt{2} r_0}{r}\left(1+ \frac{r(\cos\theta-1)}{R} \right)\,,
\eeqa
\begin{equation}\label{gzz}
g_{zz} = 1 + \frac{r_0}{r}\left(1+\frac{2r}{R}\cos\theta\right)
+ \frac{2r}{R}\cos\theta\,,
\end{equation}
\begin{equation}
g_{rr} =  1+\frac{r_0}{r}
    \left(1-\frac{2r\cos\theta}{R}\right)-
    \frac{2r\cos\theta}{R}\,,
\end{equation}
\begin{equation}
g_{\theta\theta} = r^2 \left[
    1+\frac{r_0}{r}
    \left(1-\frac{2r\cos\theta}{R}\right)-
    \frac{2r\cos\theta}{R}
    \right]\,,
\end{equation}
\begin{equation}\label{gphiphi}
g_{\phi \phi} = r^2 \sin^2 \theta \left[
    1+\frac{r_0}{r}
    \left(1-\frac{2r\cos\theta}{R}\right)-
    \frac{2r\cos\theta}{R}
    \right]\,.
\end{equation}
\end{subequations}
This solution must now be compared to the one we obtained, also in
transverse gauge, in section \ref{sec:equil},
eqs.~\reef{soltran} with $n=1$, where we solved the
equations directly in the overlap zone. That solution, in fact, provided
the basis for our claim that only perturbations of dipole type affect
the near-horizon geometry. Hence, the origin of the discrepancy must be
visible there. Indeed, comparing the solutions we find that all terms
in \reef{soltran} with $n=1$ agree
with \reef{gmunu2} except for $g_{tz}$: in
\reef{gtz} there is a term $\propto
r_0/R$ of monopole type, \ie independent of $\theta$, which is absent in
\reef{soltran2}.

The difference between \reef{soltran2} and \reef{gtz} is
pure gauge, since we can obtain the latter from the former by making
\beq\label{gau}
t\to t-\frac{\sqrt{2}\,r_0}{R}z\,,
\eeq
which is indeed a gauge transformation involving a monopole term.
However, this
does not mean that the difference is physically irrelevant. The
discrepant monopole term at order $1/R$ shows up as a constant in $A$ in
\reef{PhiAover}, and so it can be traced back to the choice of
integration constant in \reef{PhiA}. This was chosen to make $A$ vanish
at the $\psi$-rotation axis $r_2=0$, as is required for the one-form $A
d\psi$ to be well-defined. So this condition removes the freedom to change
$t\to t+ k \psi=t+ \frac{k}{R} z$, although this is not evident when
working in the overlap zone. In other words, regularity
at the rotation axis demands that we change \reef{soltran2} (for $n=1$) to the
regular gauge \reef{gau}. As a consequence, the monopole perturbation in
\reef{gtz}, even if gauge, is physical. This is an example of how
a gauge degree of freedom becomes physical due to boundary conditions.

Since this perturbation changes the monopolar part of the solution, it
induces corrections of order $1/R$ in the physical magnitudes
$\mathcal{A}$, $M$, $J$, $\kappa$, $\Omega$, with the correct values to
reproduce the exact results to linear order in $r_0/R$.
Such an effect, however, is absent in six or more dimensions, where the field
$A$ falls faster away from the source.

\subsection{The near-horizon solution}

The analysis in sec.~\ref{sec:nearhor} simplifies considerably in the
case $n=1$ and allows us to be fully explicit, since
\beq
u_1=3r-2r_0\,,\qquad u_2=r_0\,,
\eeq
and $A_1=-4/3$, $A_2=-14/3$, $B_1=-2/3$, $B_2=-4/3$. Then
\beq
{\sf A}(r)=-4r-2r_0\,,\qquad {\sf B}(r)=-2r\,,
\eeq
and
\beq\label{FGp}
{\sf F}(r)=4r+4r_0+\frac{12 r^2}{r_0}-\frac{4r^2}{r-r_0}\,,\qquad
{\sf G}'(r)= -2+\frac{8r}{r-r_0}+\frac{12 r}{r_0}\,.
\eeq
So
\beq
 a(r)=-4r-2r_0+2c(r)\,,\qquad b(r)=-2r+c(r)\,,
\eeq
and
\begin{subequations}\label{fgp}
\beqa
f(r)&=&4r+4r_0+\frac{12 r^2}{r_0}-\frac{4r^2}{r-r_0}+\frac{r c(r)}{r-r_0}
-\frac{4r}{r_0}c(r)-\frac{2r^2}{r_0}c'(r)\,,\\
g'(r)&=&-2+\frac{8r}{r-r_0}+\frac{12 r}{r_0}-\frac{2 c(r)}{r-r_0}-
\frac{4c(r)}{r_0}-\frac{2r}{r_0}c'(r)\,.
\eeqa
\end{subequations}
Boundary
conditions at the horizon and at large $r$ require that the function
$c(r)$ takes the form
\beq
c(r)=2r+r_0+\frac{r_0^2}{r}+\frac{r_0^3}{r^2}\,\tilde c(r)\,,
\eeq
where the gauge-dependent function $\tilde c(r)$ is an analytic
function of $1/r$ with $\tilde
c(r_0)=0$. The value $c(r_0)=4r_0$ guarantees that $f$ and $g'$ are
regular at $r=r_0$ in spite of the fact that
${\sf F}$ and ${\sf G}'$ are singular there.

In order to make a choice of gauge, let us impose $g(r)=f(r)$. Then
\reef{fgp} require that $\tilde c$ solves
\beq
2r^2(r-r_0)^2\tilde c''(r)-r(r-r_0)(2r-r_0)\tilde c'(r)+r_0^2\; \tilde
c(r)=0\,,
\eeq
which can be transformed into an associated Legendre equation.
The only solution with the prescribed behavior at the boundaries is
\beq
\tilde c(r)=0\,.
\eeq
Then
\beq
f(r)=g(r)=-2r+r_0\,.
\eeq
The solution in \cite{Emparan:2001wn}, to leading order in $r_0/R$, is
in fact in
this gauge.

Finally, in order to illustrate the fact that the distortion of the $S^2$ at the
horizon, $g(r_0)$, is gauge-dependent and so can take any arbitrary
value (at least to leading order in $1/R$), choose instead
\beq
\tilde c(r)=c_1\left(1-\frac{r_0}{r}\right)
\eeq
with constant $c_1$ (of course this makes $f\neq g$ in general).
$g'$ is easily integrated, and the asymptotic
behavior $g(r)+2r\to 0$ at $r\to \infty$ fixes the integration constant.
Then we find that $g(r_0)=(2c_1-1)r_0$. So the sign of the distortion of
the $S^2$ at the horizon is arbitrary, and it can even remain perfectly
round.

\section{KK phases on $\T^2$ from phases on $S^1$}
\label{app:torusphases}

\newcommand{\mt}{\mathfrak{t}}

In this appendix we show how to translate
the known results for KK  black holes on the circle
($\ie$ on $\CM^{n+2} \times S^1$) to the corresponding results for
KK black holes on the torus ($\ie$ on $\CM^{n+2} \times \T^2$),
which are used in section \ref{sec:complete}.

\subsubsection*{New dimensionless quantities}

We first recall that for KK black holes in $D$ dimensions,
the typical dimensionless quantities that are used are the dimensionless
mass $\mu$,
temperature $\mt$, and entropy $\ms$
defined by
\begin{equation}
\mu = \frac{16\pi G}{L^{D-3}}\; M \spa \ms = \frac{16 \pi
G}{L^{D-2}}\; S \spa \label{tsneut} \mt = L T  \ .
\end{equation}
These quantities were originally
introduced in \cite{Harmark:2003dg,Harmark:2004ws} for black holes on
a KK circle of circumference $L$, but we may
similarly use these definitions for KK black holes in $D$ dimensions
with a square torus of side lengths $L$.

Instead of these, we now introduce the following new dimensionless quantities,
more suitable for the application in this paper, by defining
\begin{equation}
\label{elladef}
\ell  = \mu^{-\frac{1}{D-3}} \spa a_H  =  \mu^{-\frac{D-2}{D-3}} \ms \spa
\mt_H = \mu^{\frac{1}{D-3} } \mt\,.
\end{equation}
In the KK black hole literature, entropy plots are typically given as $\ms (\mu)$. Instead
of these we then use \eqref{elladef} to
consider the area function $a_H (\ell)$, which is obtained as
\begin{equation}
\label{afroms}
a_H (\ell) = \ell^{D-2} \ms (\ell^{-D+3})\,.
\end{equation}

\subsubsection*{Map from circle to torus compactification}

In the following we use hatted quantities to refer to KK black holes
on $\CM^{n+1} \times S^1$ and unhatted quantities when referring to KK black
holes on $\CM^{n+2} \times \T^2$. Note that in the latter case
the definitions \eqref{elladef} for $\ell$, $a_H$ reduce to those in \eqref{eladef},
and that the definition
of $\mt_H$ is up to constants identical to the one in \eqref{otdef}.

Suppose we are given an entropy function $\hat \ms(\hat \mu)$ for a phase of KK
black holes on $\CM^{n+2} \times S^1$.
Any such phase lifts trivially to a phase of KK black holes on
$\CM^{n+2} \times \T^2$ that is uniform in one of the torus directions.
We want to know how we get the function $a_H(l)$ for the latter in terms of
$\hat \ms (\hat \mu)$ of the former.
It is not difficult to see that in terms of our original dimensionless quantities
we have the simple mapping
\begin{equation}
\label{msmap} \mu = \hat \mu \spa \ms = \hat \ms \spa \mt = \hat \mt
\,.
\end{equation}
It then follows from \eqref{elladef} and \eqref{afroms} that the
area function $a_H (\ell)$ of KK black holes on $\CM^{n+2} \times \T^2$ is
obtained via the mapping relation
\begin{equation}
\label{aHmap}
a_H (\ell) = \ell^{n+2} \hat \ms (\ell^{-n-1} )\,.
\end{equation}

\subsubsection*{Application to known phases}

Using now the entropy function of the uniform black string in  $\CM^{n+2} \times S^1$
\begin{equation}
\hat \ms_{\rm uni} (\hat \mu) \sim \hat \mu^{ \frac{n}{n-1}}
\end{equation}
we get from \eqref{aHmap} the result \eqref{aHmem} for $a_H^{\rm ubm} (\ell) $
of the uniform black membrane.
Furthermore, using that for small $\mu$ (or equivalently large $\ell $)
the entropy of the localized black hole in $\CM^{n+2} \times S^1$ is
\begin{equation}
\hat \ms_{\rm loc} (\hat \mu) \sim \hat \mu^{ \frac{n+1}{n}}
\end{equation}
we find via the map \eqref{aHmap} the result \eqref{aHstr} for $a_H^{\rm lbs} (\ell)$
of the localized black string in the large $\ell$ limit.

For the non-uniform string in $\CM^{n+2} \times S^1$ dimensions we know that
(see e.g. Eq.(3.16) in \cite{Harmark:2007md})
\begin{equation}
\label{snuni}
\frac{\hat \ms_{\rm nu} ( \hat \mu )}{\hat \ms_{\rm uni}  ( \hat \mu )}
= 1 - \frac{n^2}{2(n+1)(n-1)^2} \frac{\gamma_{n+2} }{\hat \mu_{{\rm GL},n+2}}
(\hat \mu-\hat \mu_{{\rm GL},n+2})^2 + \CO ( (\hat \mu - \hat \mu_{{\rm GL},n+2})^3 ) \ ,
\end{equation}
where $\hat \ms_{\rm uni}  ( \hat \mu )$ is the entropy of the uniform black string
and $\hat \mu_{{\rm GL},d}$, $\gamma_d$ can e.g. be found in Tables 2,3 of the
review \cite{Harmark:2007md}.
The result \eqref{ahnum} for the black membrane on $\CM^{n+2} \times \T^2$
with non-uniformity in one direction then follows again using
\eqref{snuni} and the map \eqref{aHmap}.

It is also known that the localized black hole and non-uniform black string phase
on $\CM^{n+2} \times S^1$ have 'copied' phases with multiple non-uniformity or
multiple localized black objects \cite{Horowitz:2002dc,Harmark:2003eg}. In particular,
 the copies (denoted with a tilde) are obtained from the original phases using
\begin{equation}
\label{coptrans} \tilde{\hat \mu} = \frac{\hat \mu}{k^{n-1}} \spa
\tilde{\hat \ms} = \frac{\hat \ms}{k^{n}}    \spa \tilde{\hat \mt} =
k \hat \mt  \ ,
\end{equation}
where $k$ is a positive integer denoting how many times the solution is copied.
Using the relation \eqref{msmap} and the definitions \eqref{elladef} then shows
that the corresponding copied phases of KK black holes on the torus obey
the transformation rule \eqref{copy}.

Finally, we give the explicit mapping used to convert the known results for
KK black holes on $\CM^5 \times S^1$ to obtain the phase diagram in Fig.~\ref{fig:KKphases7}
of KK black holes on $\CM^5 \times \T^2$. We start with the six-dimensional numerical
data that are known for the non-uniform and localized phase \cite{Kudoh:2004hs},
which have been converted to plots of points $(\hat \mu , \hat \ms)$ (see e.g.
\cite{Harmark:2007md}). Using \eqref{msmap}, \eqref{elladef} these data
are then easily transformed to curves in an $a_H (\ell)$ phase diagram
for phases in seven dimensions with a torus, according to
\begin{equation}
(\ell , a_H ) = ( \hat \mu^{-1/4}, \hat \mu^{-5/4} \hat \ms)\,.
\end{equation}

\end{appendix}

\newpage



\providecommand{\href}[2]{#2}\begingroup\raggedright\endgroup

\end{document}